\documentclass[10pt,twocolumn]{article}
\usepackage{amsmath}
\usepackage{amssymb}
\usepackage{epsfig}

\begin{document}

\setlength{\baselineskip}{12pt}

\title{Energy-aware Application Scaling on a Cloud}
\author{Ashkan Paya and Dan C. Marinescu \\
Computer Science Division \\
Department of Electrical Engineering and Computer Science \\
University of Central Florida, Orlando, FL 32816, USA \\
Email:ashkan\_paya@knights.ucf.edu, dcm@cs.ucf.edu}

\maketitle

\begin{abstract}
Cloud elasticity - the ability to use as much resources as needed at any given time - and low cost - a user pays only for the resources it consumes - represent solid incentives for many organizations to migrate some of their computational activities to a public cloud. As the interest in cloud computing grows, so does the size of the cloud computing centers and their energy footprint. The realization that power consumption of cloud computing centers is significant and it is expected to increase substantially in the future motivates our interest in scheduling and scaling algorithms which minimize power consumption. We propose energy-aware application scaling and resource management algorithms. Though targeting primarily  the {\it Infrastructure as a Service (IaaS)}, the system models and the algorithms we propose can be applied to the other cloud delivery models and to private clouds.
\end{abstract}


\section{Introduction}
\label{Introduction}
\medskip

The power consumption of large-scale data centers and their costs for energy and for cooling are significant and are expected to increase in the future. In $2006$, the $6\,000$ data centers in the U.S. reportedly consumed $61 \times 10^{9}$ KWh of energy, $1.5\%$ of all electricity consumption in the country, at a cost of $\$4.5$ billion \cite{Vrbsky10}. The power consumption of data centers and the network infrastructure is predicted to reach $10,300$ TWh/year\footnote{One TWh (Tera Watt Hour) is equal to $10^{12}$ Wh.} in $2030$, based on $2010$ levels of efficiency \cite{Preist10}. These increases are expected in spite of the extraordinary reduction in  energy requirements for computing activities; over the past $30$ years the energy efficiency per transistor on a chip has improved by six orders of magnitude.
Energy-aware scheduling and scaling algorithms could reduce the energy consumption of cloud computing centers at a time when the interest in cloud computing is on the raise.

Scaling is the process of allocating additional resources to a cloud application in response to a request consistent with the {\it Service Level Agreement (SLA} between a cloud service provider (CSP) and a  cloud user. We distinguish two scaling modes, horizontal and vertical scaling. {\it Horizontal scaling} is the most common mode of scaling on a cloud; it is done by increasing the number of Virtual Machines (VMs) when the load of application $\mathcal{A}$ increases and reducing this number when the load decreases. Often, this leads to an increase of communication bandwidth consumed by the application. Load balancing among the running VMs is critical for this mode of operation. For a very large application multiple load balancers may need to cooperate with one another. In some instances the load balancing is done by a frontend server which distributes incoming requests of a transaction-oriented system to backend servers. {\it Vertical scaling} keeps the number of VMs of an application constant, but increases the amount of resources allocated to each one of them. This can be done either by migrating the VMs to more powerful servers, or by keeping the VMs on the same servers, but increasing their share of the CPU time. The first alternative involves additional overhead; the VM is stopped, a snapshot of it is taken,  the file is transported to a more powerful server, and the VM is restated at the new site.

We assume a clustered organization of a cloud similar to the one described in \cite{Paya13} and consider three levels of resource allocation decision making: (a) the local system which has accurate information about its state; (b) the cluster leader which has less accurate information about the servers in the cluster; and (c) global decisions involving multiple clusters. The basic philosophy of our approach is to define a power-optimal operation region for each server and to attempt to maximize the number of servers operating in this region.

In a cloud environment we should recast some of the traditional policy objectives of a distributed system; for example, load balancing should be weighted against power consumption, it may be beneficial to turnoff or switch to a sleep state lightly loaded servers to save energy. We also believe that the ability of a cloud infrastructure to respond to a scaling request should be considered as an important element of the Quality of Service provided by the cloud.

In Section \ref{Motivation} we overview the energy consumption of cloud computing centers and of individual servers. Then, in Section \ref{CloudResourceManagement} we discuss the mechanisms for the implementation of resource management policies: admission control, capacity allocation, load balancing, energy optimization, and QoS guarantees; virtually none of the mechanisms proposed so far in the literature integrate solutions for all five classes of policies.

We develop an energy-aware operation model for a server $\mathcal{S}$  which identifies an optimal operating region, two suboptimal, and two undesirable ones. The model described in Section \ref{SystemModel} is then used for the algorithms introduced in Section \ref{ScalingAlgorithms}. The algorithms we propose balance minimization of energy consumption and the ability to respond to scaling requests; at the same time, they implement a load balancing in the sense described above and can also be used for admission control and capacity allocation in systems where the servers are clustered together in clusters of moderate size. In Sections \ref{SimulationExperiment} we discuss a simulation experiment and then in Section \ref{Conclusions} we review our results and discuss future work.

\section{Motivation and Related Work}
\label{Motivation}
\medskip

The power consumption of servers has increased over time. Table \ref{ServerEnergyConsumption}  \cite{Koomey07} shows the evolution of the average power consumption for volume (Vol) servers - servers with a price less than \$ 25 K, mid-range (Mid) servers - servers with a price between \$25 K and \$499 K, and high-end (High) servers - servers with a price tag larger than \$500 K.

\begin{table*}[!ht]
\begin{center}
\caption{Estimated average power use of volume, mid-range, and high-end servers (in Watts) along the years \cite{Koomey07}.}
\label{ServerEnergyConsumption}
\begin{tabular} {|c|ccccccc|}
\hline
Type   & 2000  & 2001  & 2002  & 2003  & 2004  & 2005  & 2006 \\
\hline
Vol    & 186   & 193   & 200   & 207   & 213   & 219   & 225 \\
Mid    & 424   & 457   & 491   & 524   & 574   & 625   & 675 \\
High   & 5,534 & 5,832 & 6,130 & 6,428 & 6,973 & 7,651 & 8,163 \\
 \hline
\end{tabular}
\end{center}
\end{table*}

The largest consumer of power of a system is the processor, followed by memory, and storage systems. The power consumption can vary from 45W to 200W per multi-core CPU; newer processors include power saving technologies. Large servers often use $32-64$ {\it Dual In-line Memory Modules (DIMMs)}; the power consumption of one DIMM is in the $5 - 21$ W range.  Server secondary memory cooling requires additional power; a server with $2-4$ {\it Hard Disk Drives (HDDs)} consumes $24 - 48$ W.

A strategy to reduce energy consumption by disk drives is to concentrate the workload on a small number of disks and allow the others to operate in a low-power mode. One of the techniques to accomplish this  is based on replication. A replication strategy based on a sliding window is reported in \cite{Vrbsky10};  measurement results indicate that it performs better than  LRU, MRU, and LFU\footnote{LRU (Least Recently Used), MRU (Most Recently Used), and LFU(Least Frequently Used) are replacement policies used by memory hierarchies  for caching and paging.} policies for a range of file sizes, file availability, and number of client nodes and the power requirement is reduced by as much as $31\%$.

Another technique is based on data migration. The system in \cite{Hasebe10} uses data storage in virtual nodes managed with a distributed hash table; the migration is controlled by two algorithms,  a short-term optimization algorithm used for gathering or spreading virtual nodes according to the daily variation of the workload so that the number of active physical nodes is reduced to a minimum, and a  long-term optimization algorithm, used for coping with changes in the popularity of data over a longer period, e.g., a week.

\begin{figure}[!ht]
\begin{center}
\includegraphics[width=7.5cm]{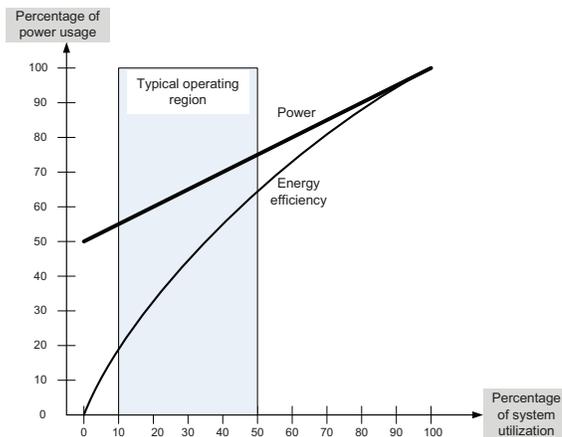}
\end{center}
\caption{Even when power requirements scale linearly with the load, the energy efficiency of a computing system is not a linear function of the load; even when idle, a system may use $50\%$ of the power corresponding to the full load. Data collected over a long period of time shows that the typical operating region for data center servers is in the range $10\% - 50\%$ of the load. }
\label{EnergyEfficiency}
\end{figure}

In an ideal world, the energy consumed by an idle system should be near zero and grow linearly with the system load. In real life, even systems whose power requirements scale linearly,  when idle use more than half the power they use at full load, see Figure \ref{EnergyEfficiency} \cite{Barroso07}.

The operating efficiency of a system is captured by an expression of ``performance per Watt of power.'' It is widely reported that during the last two decades the performance of computing systems has increased much faster than their operating efficiency; for example, during the period $1998$ till $2007$, the performance of supercomputers has increased $7,000\%$  while their operating efficiency has increased only $2,000\%$.

Energy-proportional systems could lead to large savings in energy costs for computing clouds.
An {\it energy-proportional} system consumes no power when idle, very little power under a light load and, gradually, more power as the load increases. By definition, an ideal energy-proportional system is always operating at $100\%$ efficiency. Humans are a good approximation of an ideal energy proportional system; the human energy consumption is about $70$ W at rest, $120$ W on average on a daily basis, and can go as high as $1,000 - 2,000$ W during a strenuous, short time effort \cite{Barroso07}.

Different subsystems of a computing system behave differently in terms of energy efficiency;
while many processors have reasonably good energy-proportional profiles,  significant improvements in memory and disk subsystems are necessary. The processors used in servers consume less than one-third of their peak power at very-low load and have a dynamic range\footnote{The dynamic range in this context is the difference between the upper and the lower limits of the power consumption of the device function of the load placed on the device. A large dynamic range means that the device is better, it is able to operate at a lower fraction of its peak power when its load is low.} of more than $70\%$ of peak power; the processors used in mobile and/or embedded applications are better in this respect. According to \cite{Barroso07} the dynamic power range of other components of a system  is much narrower: less than $50\%$ for DRAM, $25\%$ for disk drives, and $15\%$ for networking switches.

A number of proposals have emerged for {\it energy proportional} networks; the energy consumed by such networks is proportional with the communication load. For example, in \cite{Abts10} the authors argue that a data center network based on a flattened butterfly topology is more energy and cost efficient. High-speed channels typically consist of multiple serial lanes with the same data rate;  a physical unit is stripped across all the active lanes. Channels commonly operate plesiochronously\footnote{Different parts of the system are almost, but not quite perfectly, synchronized; in this case, the core logic in the router operates at a frequency different from  that of the I/O channels.} and are always on, regardless of the load, because they must still send idle packets to maintain byte and line alignment across the multiple lines. An example of an energy proportional network is {\it InfiniBand}.

\section{Cloud Resource Management}
\label{CloudResourceManagement}
\medskip

Cloud resource management policies can be loosely grouped into five classes: (a)
Admission control; (b) Capacity allocation; (c) Load balancing; (d) Energy optimization; and (e) Quality of service (QoS) guarantees.

The explicit goal of an admission control policy is to prevent the system from accepting workload in violation of high-level system policies; a system should not accept additional  workload  preventing it from completing work already in progress or contracted. Limiting the workload requires some knowledge of the global state of the system; in a dynamic system such knowledge, when available, is at best obsolete.  Capacity allocation means to allocate resources for individual instances; an instance is an activation of a service.  Assigning instances to physical servers is subject to multiple global optimization constraints and requires a search in a very large search space; moreover the state of individual systems changes rapidly.

Load balancing and energy optimization can be done locally, but global load balancing and energy optimization policies encounter the same difficulties as the one we have already discussed. Load balancing and energy optimization are correlated and affect the cost of providing the services. Indeed, it was predicted that by 2012 up to $40\%$ of the budget of IT enterprise infrastructure would be spent on energy \cite{Dikaiakos09}.

The common meaning of the term ``load balancing'' is that of evenly distributing the system load to a set of servers. For example, consider the case of four identical servers, $A, B, C$ and $D$ whose relative loads are $70\%, 50\%, 30\%$ and $10\%$, respectively, of their capacity; as a result of a perfect load balancing all servers would end with the same load, $40\%$ of each server's capacity. 

In cloud computing a  critical goal is minimizing the cost of providing the service and, in particular, minimizing the energy consumption. This leads to a different meaning of the term ``load balancing;'' instead of having the load evenly distributed amongst all servers, we wish to concentrate it and use the smallest number of servers while switching the others to a standby mode, a state where a server uses very little energy. In our example, the load from $D$ will migrate to $A$ and the load from $C$ will migrate to $B$; thus, $A$ and $B$ will be loaded at $80\%$ of their capacity, while $C$ and $D$ will be switched to standby mode.   

{\it Quality of Service (QoS)} is that aspect of resource management probably the most difficult to address and, at the same time, possibly the most critical for the future of cloud computing. For applications with hard or soft deadlines it means the ability to complete a task before its deadline.   For many applications  QoS  requires also the ability to scale.

Often, resource management strategies jointly target the performance and the power consumption. The {\it Dynamic Voltage and Frequency Scaling (DVFS)}\footnote{Dynamic voltage and frequency scaling is a power management technique to increase or decrease the operating voltage or frequency of a processor to increase the instruction execution rate and, respectively, to  reduce the amount of heat generated and to conserve power.} techniques such as Intel's {\it SpeedStep} and AMD's {\it PowerNow} lower the voltage and the frequency to decrease the power consumption\footnote{The power consumption $P$ of a CMOS-based circuit is: $P=\alpha \cdot C_{eff} \cdot V^{2} \cdot f$ with: $\alpha$ - the switching factor, $C_{eff}$ - the effective capacitance, $V$ - the operating voltage, and $f$ - the operating frequency.}. Motivated initially by the need to save power for mobile devices, these techniques have migrated virtually to all processors including the ones used for high performance servers.

As a result of lower voltages and frequencies, the performance of processors decreases, but at a substantially slower rate \cite{LeSueur10}. Table \ref{DVFS} shows the dependence of the normalized performance and the normalized energy consumption of a typical modern processor on the clock rate; as we can see, at $1.8$ GHz we save $18\%$ of the energy required for maximum performance, while the performance is only $5\%$ lower than the peak performance, achieved at $2.2$ GHz. This seems a reasonable energy-performance tradeoff!

\begin{table}[!ht]
\begin{center}
\caption{The normalized performance and energy consumption, function of the processor speed; the performance decreases at a lower rate than does the energy when the clock rate decreases \cite{LeSueur10}. }
\label{DVFS}
\begin{tabular} {|c|c|c|}
\hline
Speed    & Energy & Performance \\
 (GHz)    &    ($\%$)           &      ($\%$)                \\
 \hline \hline
  0.6    &     0.44           &      0.61    \\
  0.8    &     0.48           &      0.70    \\
  1.0    &     0.52           &      0.79    \\
  1.2    &     0.58           &      0.81    \\
  1.4    &     0.62           &      0.88    \\
  1.6    &     0.70           &      0.90    \\
  1.8    &     0.82           &      0.95    \\
  2.0    &     0.90           &      0.99    \\
  2.2    &     1.00           &      1.00    \\
 \hline
\end{tabular}
\end{center}
\end{table}

Virtually all optimal, or near-optimal, mechanisms to address the five classes of policies do not scale up and typically target a single aspect of resource management, e.g., admission control, but ignore energy conservation; many require complex computations that cannot be done effectively in the time available to respond. The performance models are very complex, analytical solutions are intractable, and the monitoring systems used to gather state information for these models can be too intrusive and unable to provide accurate data. Many techniques are concentrated on system performance in terms of throughput and time in system, but they rarely include energy trade-offs or QoS guarantees. Some techniques are based on unrealistic assumptions; for example, capacity allocation is viewed as an optimization problem,  but under the assumption that servers are protected from overload.

The effort to reduce power consumption covers computing, networking, and storage activities of a data center. A $2010$ report shows that a typical Google cluster spends most of its time within the $10-50\%$ CPU utilization range; there is a mismatch between server workload profile and server energy efficiency \cite{Abts10}. A similar behavior is also seen in the data center networks; these networks operate in a very narrow dynamic range,  the power consumed when the network is idle is  significant compared to the power consumed when the network is fully utilized.

Many proposals argue that dynamic resource provisioning is necessary to minimize power consumption. Two main issues are critical for energy saving: the amount of resources allocated to each application and the placement of individual workloads. For example, a resource management framework combining a utility-based dynamic Virtual Machine provisioning manager with a dynamic VM placement manager to minimize power consumption and reduce Service Level Agreement violations is presented in \cite{Van10}.

\section{The System Model}
\label{SystemModel}
\medskip

{\bf Notations.} Table \ref{Notations} contains a summary of notations describing the cluster, the servers in the cluster, the applications, and the parameters of various algorithms. As a general rule we use
calligraphic upper-case characters as names; e.g., $\mathcal{S}_{k}$ - server $k$ and $\mathcal{O}_{k}$ - optimal operating region; lower-case Greek alphabet for constants, e.g., $\alpha_{k}^{opt,low}$ - performance at the lower boundary of the optimal region and $\lambda_{i}$ the largest rate of increase in demand for CPU cycles of application $\mathcal{A}_{i}$; and lower-case Latin alphabet for variables, e.g., $a_{k}(t)$ - the demand for CPU cycles by all applications running on server $\mathcal{S}_{k}$ at time $t$.

\begin{table*}[ht!]
\begin{center}
\caption{The notations used to describe the model.}
\label{Notations}
\begin{tabular} {|lll|}
\hline
$\mathcal{C}$            & $\rightarrow$ & one of the clusters of the cloud \\
$n_{\mathcal{C}}$        & $\rightarrow$ & number of servers in $\mathcal{C}$ \\
$n_{\mathcal{C}}^{r}(t)$ & $\rightarrow$ & number of active/running servers in $\mathcal{C}$ at time $t$ \\
$n_{\mathcal{C}}^{i}(t)$ & $\rightarrow$ & number of idle servers in $\mathcal{C}$ at time $t$ \\
$n_{\mathcal{C}}^{s}(t)$ & $\rightarrow$ & number of servers in $\mathcal{C}$ in sleep mode at time $t$ \\
$n_{\mathcal{C}}^{h}(t)$ & $\rightarrow$ & number of servers in $\mathcal{C}$ in hibernate mode at time $t$ \\
$\mathcal{L}_{\mathcal{C}}$ & $\rightarrow$  & leader of $\mathcal{C}$ \\
$\tau^{i}$    & $\rightarrow$            & in-cluster communication period   \\
$\tau^{c}$    & $\rightarrow$            & intra-cluster communication period   \\
\hline
$\mathcal{S}_{k}$        & $\rightarrow$ & server in $\mathcal{C}$ \\
$\gamma_{k}$               & $\rightarrow$ & constant quantifying the highest level of performance $\mathcal{S}_{k}$ can deliver \\
$\tau_{k}$ & $\rightarrow$ &  $\mathcal{S}_{k}$ reallocation interval    \\
$\beta_{k}^{idle}$         & $\rightarrow$ & $\mathcal{S}_{k}$ energy consumption when idle \\
$\beta_{k}^{sleep}$        & $\rightarrow$ & $\mathcal{S}_{k}$ energy consumption when in stand-by/sleep state \\
$f_{k}$                    & $\rightarrow$ & $\mathcal{S}_{k}$ performance versus energy characteristic function \\
$a_{k}(t)$                 & $\rightarrow$ & demand for CPU cycles of all applications running on $\mathcal{S}_{k}$ at time $t$ \\
$b_{k}(t)$               & $\rightarrow$ & power consumed by $\mathcal{S}_{k}$ at time $t$\\
$\beta_{k}^{opt,low}$    & $\rightarrow$ & $\mathcal{S}_{k}$  energy consumption at the low boundary of the optimal region \\
$\alpha_{k}^{opt,low}$   & $\rightarrow$ & $\mathcal{S}_{k}$ normalized performance at low boundary of optimal region \\
$\beta_{k}^{opt,high}$   & $\rightarrow$ & $\mathcal{S}_{k}$ energy consumption at the high boundary of optimal region\\
$\alpha_{k}^{opt,high}$  & $\rightarrow$ & $\mathcal{S}_{k}$ normalized performance at high boundary of optimal region \\
$\beta_{k}^{sopt,low}$   & $\rightarrow$ & $\mathcal{S}_{k}$  energy consumption at low boundary of lower suboptimal region\\
$\alpha_{k}^{sopt,low}$  & $\rightarrow$ & $\mathcal{S}_{k}$ normalized performance at low boundary of lower suboptimal region\\
$\beta_{k}^{sopt,high}$  & $\rightarrow$ & $\mathcal{S}_{k}$ energy consumption at high boundary of upper suboptimal region \\
$\alpha_{k}^{sopt,high}$ & $\rightarrow$ & $\mathcal{S}_{k}$ normalized performance at high boundary of upper suboptimal region \\
$\beta_{k}^{0}$          & $\rightarrow$ & $\mathcal{S}_{k}$  energy consumption when idle \\
 \\
\hline
$\mathcal{A}_{i,k}$      & $\rightarrow$ & application running on $\mathcal{S}_{k}$ \\
$a_{i,k}(t)$             & $\rightarrow$ & demand for CPU cycles of application $\mathcal{A}_{i,k}$ at time $t$ \\
$c_{i,k}(t)$             & $\rightarrow$ & change in demand for CPU cycles of application $\mathcal{A}_{i}$ at time $t$ \\
$\lambda_{i,k}$            & $\rightarrow$ & highest rate of increase in demand for CPU cycles of application $\mathcal{A}_{i,k}$ \\
$p_{i,k}(t)$             & $\rightarrow$ & cost of migrating application $\mathcal{A}_{i,k}$ at time $t$ \\
$q_{i,k}(t)$             & $\rightarrow$ & cost of horizontal scaling of application $\mathcal{A}_{i,k}$ at time $t$ \\
\hline
\end{tabular}
\end{center}
\end{table*}

{\bf Clustered organization.} We assume that the cloud storage and computational servers are partitioned into $N$ clusters. A self-organization algorithm for clustering based on a biased random walk is introduced in  \cite{Paya13}; a cluster $\mathcal{C}$ has a leader, $\mathcal{L}_{\mathcal{C}}$\footnote{In \cite{Paya13} the leader is called a {\it core} node,
it is self appointed based on the degree of its connectivity.}, which maintains relatively accurate information about the free capacity of individual servers in the cluster and communicates with the leaders of the other clusters for the implementation of global resource management policies.

An advantage of a clustered organization is that a large percentage of scheduling decisions are based on local, therefore more accurate, information. The servers in the cluster report to the leader the current load and other relevant state information every $\tau^{i}$ units of time, or earlier if the need to migrate an application is anticipated. The cluster leader  can thus implement the resource management policies discussed in Section \ref{CloudResourceManagement} more effectively. In this paper we are only concerned with {\it in-cluster scheduling} coordinated by $\mathcal{L}_{\mathcal{C}}$, the leader. {\it Inter-cluster scheduling} is based on less accurate information as the leader $\mathcal{L}_{\mathcal{C}}$ exchanges information with other leaders less frequently, every $\tau^{l} >> \tau^{c}$ units of time.

We assume that the scheduler of the {\it Virtual Machine Monitor (VMM)/hypervisor} of server $\mathcal{S}_{k}$ runs the {\it Borrowed Virtual Time (BVT)} scheduling algorithm and interacts with the {\it Server Application Management (SAM)} component of the VMM discussed in this paper to ensure that, once an application is allocated the requested amount of CPU cycles, the QoS requirements of the application are satisfied.  The objective of the BVT algorithm  is to support low-latency dispatching of real-time applications, as well as a weighted sharing of the CPU among several classes of applications \cite{Duda99}; it supports scheduling of a mix of applications, some with hard, some with soft real-time constraints, and applications demanding only a best-effort.

{\bf The servers.} Amazon Web Services (AWS) is representative for the {\it IaaS} cloud delivery model. AWS offers several classes of services; the servers in each class are characterized by the architecture, CPU execution rate, main memory, disk space, and I/O bandwidth. The more powerful the server, the higher the cost per hour for the class of service. AWS also supports multiple costs models for reserved instances, on-demand instances, and spot instances.

To avoid complicating our model, we assume that a server $\mathcal{S}_{k}$ has a computational constant $\gamma_{k}$ which quantifies the highest level of performance it can deliver. We also assume that the actual cost for the user is captured by $\gamma_{k}$ thus, whenever feasible, an application $\mathcal{A}$ is assigned to the server $\mathcal{S}_{k}$ with the lowest $\gamma_{k}$.

$\mathcal{S}_{k}$ makes scheduling decisions every $\tau_{k}$ units of time. We assume that a server can be either running, idle, in a stand-by/sleep state, or in hibernate state. As we have seen in Sections \ref{Motivation} and \ref{CloudResourceManagement}, an {\it idle} server consumes a fair amount of energy thus, the cluster management algorithms should avoid keeping any server in this state. In the {\it sleep} state the power to non-essential components, such as primary and secondary storage, is turned off, thus  a server consumes only a small amount of energy; the cluster management  algorithms should switch an idle server to the sleep mode as soon as feasible. Some operating systems support a {\it hibernate} state; in this state, the data in physical memory is saved on the disk and the system is powered off. When the need to use the system arises the system boots up and loads from the disk the memory image saved when the system was forced to enter this state. Rebooting the system takes some time, but no power is used while the system is in the  hibernate state.

There are $n_{\mathcal{C}}$ servers in a cluster $\mathcal{C}$; at time $t$,  $n_{\mathcal{C}}^{r}(t)$ of them are in a running state, $n_{\mathcal{C}}^{s}(t)$ in a sleep state, $n_{\mathcal{C}}^{h}(t)$ in a hibernate state, and $n_{\mathcal{C}}^{i}(t)$ could be idle

\begin{equation}
n_{\mathcal{C}} = 1 + n_{\mathcal{C}}^{r}(t) + n_{\mathcal{C}}^{s}(t) + n_{\mathcal{C}}^{i}(t) + n_{\mathcal{C}}^{h}(t).
\end{equation}
We do not see a good reason to keep servers in a hibernate or an idle state thus, we assume that
\begin{equation}
n_{\mathcal{C}} = 1 + n_{\mathcal{C}}^{r}(t) + n_{\mathcal{C}}^{s}(t).
\end{equation}

\begin{figure*}[!ht]
\begin{center}
\includegraphics[width=12cm]{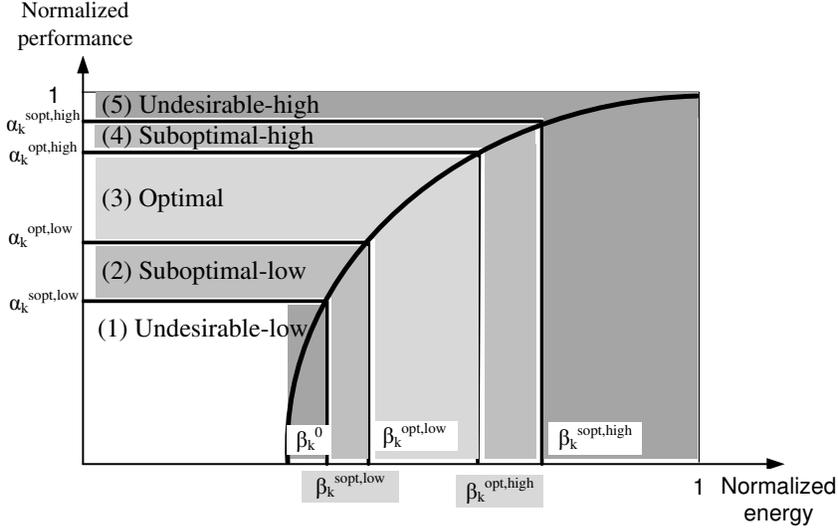}
\end{center}
\caption{Normalized performance versus normalized power consumption characteristic of  $\mathcal{S}_{k}$. There are 5 operating regions: (1) undesirable-low; (2) suboptimal-low; (3) optimal; (4) suboptimal-high; and (5) undesirable high.}
\label{ServerPerformanceVsEnergy}
\end{figure*}

The normalized performance of server $\mathcal{S}_{k}$ depends on the power level
$a_{k}(t) = f_{k} [b_{k}(t)]$. We distinguish three desirable operating regions for  $\mathcal{S}_{k}$, Figure \ref{ServerPerformanceVsEnergy}:

\begin{enumerate}
\item
$\mathcal{R}_{3}$ - optimal performance versus power consumption region
\begin{equation}
\begin{array}{c}
\beta_{k}^{opt,low}  \le b_{k}(t) \le \beta_{k}^{opt,high} \\
\alpha_{k}^{opt,low}  \le { a_{k}(t)} \le \alpha_{k}^{opt,high}.
\end{array}
\end{equation}

\item
$\mathcal{R}_{2}$ - lower  suboptimal region

\begin{equation}
\begin{array}{c}
\beta_{k}^{sopt,low}  \le b_{k}(t) \le \beta_{k}^{opt,low} \\
\alpha_{k}^{sopt,low}  \le {a_{k}(t)} \le \alpha_{k}^{opt,low}.
\end{array}
\end{equation}

\item
$\mathcal{R}_{4}$ - upper suboptimal region

\begin{equation}
\begin{array}{c}
\beta_{k}^{opt,high}  \le b_{k}(t) \le \beta_{k}^{sopt,high} \\
\alpha_{k}^{opt,high}  \le {a_{k}(t)} \le \alpha_{k}^{sopt,high}.
\end{array}
\end{equation}

\end{enumerate}

There are also two undesirable operating regions,

\begin{enumerate}
\item
$\mathcal{R}_1$ - undesirable low operating region

\begin{equation}
\begin{array}{c}
\beta_{k}^{0}  \le b_{k}(t) \le \beta_{k}^{sopt,low} \\
0  \le {a_{k}(t)} \le \alpha_{k}^{sopt,low}
\end{array}
\end{equation}

\item
$\mathcal{R}_5$ - undesirable high operating region

\begin{equation}
\begin{array}{c}
\beta_{k}^{sopt, high}  \le b_{k}(t) \le 1 \\
\alpha_{k}^{sopt,high}  \le { a_{k}(t)} \le 1
\end{array}
\end{equation}

\end{enumerate}

The {\it energy efficiency} of a system $\xi_{k}$ or as we shall call it {\it efficiency} is measured as performance per Watts of power. The  average efficiency  in the optimal region is larger than in all other regions:

\begin{equation}
\bar {\xi}^{3} > \max \left( \bar{\xi}^{1}, \bar{\xi}^{2}, \bar{\xi}^{4}, \bar{\xi}^{5}\right).
\end{equation}

The  average efficiency of servers in region $\mathcal{R}_{i}$ with $n_{\mathcal{R}_{i}}$ servers, $1 \le i \le 5$ is

\begin{equation}
\bar {\xi}^{i} = { 1 \over {n_{\mathcal{R}_{i}}}} \sum_{S_{k} \in \mathcal{R}_{i}} \xi_{k}.
\end{equation}

Then the average cluster efficiency is

\begin{equation}
\label{AverageClusterEfficiency}
\bar {\xi}_{\mathcal{C}} = {1 \over 5} \sum_{i=1}^{5} \bar {\xi}^{i} .
\end{equation}

The efficiency of server in each one of the five operating regions can be approximated as:

\begin{equation}
\xi_{k}^{1} = { \alpha_{k}^{sopt, low}  \over { \beta_{k}^{sopt, low} - \beta_{k}^{0} }},
\end{equation}

\begin{equation}
\xi_{k}^{2} = { {\alpha_{k}^{opt, low} - \alpha_{k}^{sopt, low}} \over { \beta_{k}^{opt, low} - \beta_{k}^{sopt,low} }},
\end{equation}

\begin{equation}
\xi_{k}^{3} = { {\alpha_{k}^{opt, high} - \alpha_{k}^{opt, low}} \over { \beta_{k}^{opt, high} - \beta_{k}^{opt,low} }},
\end{equation}

\begin{equation}
\xi_{k}^{4} = { {\alpha_{k}^{sopt, high} - \alpha_{k}^{opt, high}} \over { \beta_{k}^{sopt, high} - \beta_{k}^{opt,high} }},
\end{equation}

\begin{equation}
\xi_{k}^{5} = { 1 - \alpha_{k}^{sopt, high} \over { 1 - \beta_{k}^{sopt,high} }}.
\end{equation}

{\bf The leader.} The leader, $\mathcal{L}_{\mathcal{C}}$, maintains static and dynamic information about all servers in $\mathcal{C}$. Static information includes:
\begin{itemize}
\item
$\mathcal{S}_{k}, k \in (1,n_{\mathcal{C}})$ - the serverId;
\item
$\gamma_{k}$ - constant quantifying the highest level of performance of $\mathcal{S}_{k}$;
\item
$\alpha_{k}^{sopt,low}$ $\alpha_{k}^{opt,low}$, $\alpha_{k}^{opt,high}$, and $\alpha_{k}^{sopt,high}$, the normalized performance boundaries of different operating regions.
\item
$\tau_{k}$ - the reallocation interval .
\end{itemize}
The dynamic information is $a_{k}(t)$, the demand for CPU cycles of all applications running on $\mathcal{S}_{k}$ at time $t$. This information is reported periodically with period $\tau^{i}$ and whenever the server determines that it needs to migrate an application or to create additional VMs for an application. The dynamic information includes: application Id,  $\lambda_{i}$ - the largest rate of increase in demand for CPU cycles of application $\mathcal{A}_{i}$, and the parameters reflecting the cost for migration - $p_{i}(t)$ and $q_{i}(t)$. $\mathcal{L}_{\mathcal{C}}$ does not maintain information about applications. To minimize communication costs, the reporting period $\tau^{i}$ is much larger than the rescheduling period of individual clusters.

\medskip

Call $a_{i}(t)$ the CPU cycles required by application $\mathcal{A}_{i,k}$ at time $t$ and $c_{i,k}(t)$ the rate of change in demand for CPU cycles; $c_{i,k}(t)> 0 $ means $\mathcal{A}_{i,k}$ needs additional CPU cycles, while $c_{i,k}(t) < 0 $ means that the application needs less. During a reallocation cycle the increase in demand for CPU cycles of application $\mathcal{A}_{i,k}$ is limited

\begin{equation}
a_{i,k}(t+\tau_{k}) \le a_{i,k}(t) + \lambda_{i} \tau_{k}.
\end{equation}

At each reallocation instance, server $\mathcal{S}_{k}$ determines its available capacity

\begin{equation}
d_{k}(t) = \alpha_{k}^{opt,high} - {1 \over \gamma_{k}} \sum_{i} a_{i,k}(t),
\end{equation}
as well as the largest possible demand for CPU cycles at the end of that reallocation interval

\begin{equation}
g_{k}(t+ \tau_{k}) = \sum_{i} \left( a_{i,k}(t) + \lambda_{i} \tau_{k} \right).
\end{equation}
There is no need to communicate with the leader and reallocation of CPU cycles could be done locally when
\begin{equation}
\label{LocalVerticalScaling}
 \alpha_{k}^{opt,low} \le g_{k}(t+ \tau_{k})/\gamma_{k} \le  \alpha_{k}^{opt,high}.
\end{equation}

\section{Energy-aware Scaling Algorithms}
\label{ScalingAlgorithms}

The objective of the algorithms introduced in this section
is to ensure that the largest possible number of active servers
operate within the boundaries of their respective optimal operating regions.
Another critical policy is to migrate applications from a lightly
loaded server and then to switch the server to a sleep state and avoid at all
costs to keep a server in an idle state.

Some scheduling decisions are made locally by each server, others require the
intervention of the cluster leader. We assume that a server is multi-threaded and the
application management component of the VMM
can update frequently, every $\tau_{k}$ units of time, $a_{i,k}(t)$ - the current
application CPU consumption - and predict the consumption at the beginning of the next
reallocation interval.

The scaling decisions are made at several levels; they are listed in the order of their complexity:

\begin{enumerate}
\item
Local decision - carry out a vertical scaling using local resources, no need to interact with the leader.
\item
In-cluster, horizontal or vertical scaling; migrate some of the applications to other servers identified by the leader.
\item
Inter-cluster scaling; the leader, $\mathcal{L}_{\mathcal{C}}$. determines that cluster $\mathcal{C}$ does not have the available capacity to respond to a request to increase the allocation by application $\mathcal{A}$. Thus, $\mathcal{L}_{\mathcal{C}}$ must interact with the leaders of other clusters to satisfy this request. This case is not addressed in the paper.
\end{enumerate}

{\it Local, vertical scaling.} The first option of a server in response to a request to increase the CPU allocation of an application is to attempt to carry out local vertical scaling even if this leads to  operation
in a suboptimal region; to avoid the cost of application migration a server could
operate for a relatively short period of time in its $\mathcal{H}_{k}$ or $\mathcal{H}_{k}$
suboptimal regions.

\underline{Local scheduling decisions} take into account the current demand for CPU cycles as well as the maximum anticipated load at the end of the current and the next scheduling cycle. Local vertical scaling can be done if

\begin{equation}
\begin{array}{c}
g_{k}(t+ \tau_{k}) \le \gamma_{k}  a_{k}^{opt,high} \\
\text{and} \\
\gamma_{k}  a_{k}^{opt,low} \le g_{k}(t+ 2\tau_{k})\le \gamma_{k} a_{k}^{opt,high}.
\end{array}
\end{equation}

\underline{In-cluster scaling.} The server $\mathcal{S}_{k}$ sends a warning that it will operate in the upper sub-optimal region when

\begin{equation}
\begin{array}{c}
g_{k}(t+ \tau_{k}) \le \gamma_{k}  \alpha_{k}^{opt,high}\\
\text{and} \\
\gamma_{k} \alpha_{k}^{opt,high} \le g_{k}(t+ 2\tau_{k}) \le \gamma_{k}  \alpha_{k}^{sopt,high}.
\end{array}
\end{equation}

$\mathcal{S}_{k}$ identifies the application(s) which need additional VMs or have to be migrated to a more powerful server and  sends to the leader $\mathcal{L}_{\mathcal{C}}$ an imperative request for horizontal or vertical scaling when

\begin{equation}
\begin{array}{c}
g_{k}(t+ \tau_{k}) \le \gamma_{k} \alpha_{k}^{opt,high}\\
\text{and} \\
g_{k}(t+ 2\tau_{k}) > \gamma_{k} \alpha_{k}^{sopt,high}.
\end{array}
\end{equation}

The server reports to $\mathcal{L}_{\mathcal{C}}$ low future utilization and becomes a candidate for entering a sleep state when

\begin{equation}
\begin{array}{c}
g_{k} (t+ \tau_{k}) \le \gamma_{k}  \alpha_{k}^{opt,high}  \\
\text{and} \\
\gamma_{k} \alpha_{k}^{sopt,low} \le g_{k} (t+ 2\tau_{k}) \le \gamma_{k} \alpha_{k}^{opt,low}.
\end{array}
\end{equation}

Server $\mathcal{S}_{k}$ sends an imperative request to migrate the applications and
be switched to the sleep state when

\begin{equation}
\begin{array}{c}
g_{k} (t+ \tau_{k}) \le \gamma_{k} \alpha_{k}^{opt,high}  \\
\text{and} \\
g_{k} (t+ 2\tau_{k}) \le \gamma_{k} \alpha_{k}^{sopt,low}.
\end{array}
\end{equation}

\begin{figure*}[!ht]
\begin{center}
\includegraphics[width=12cm]{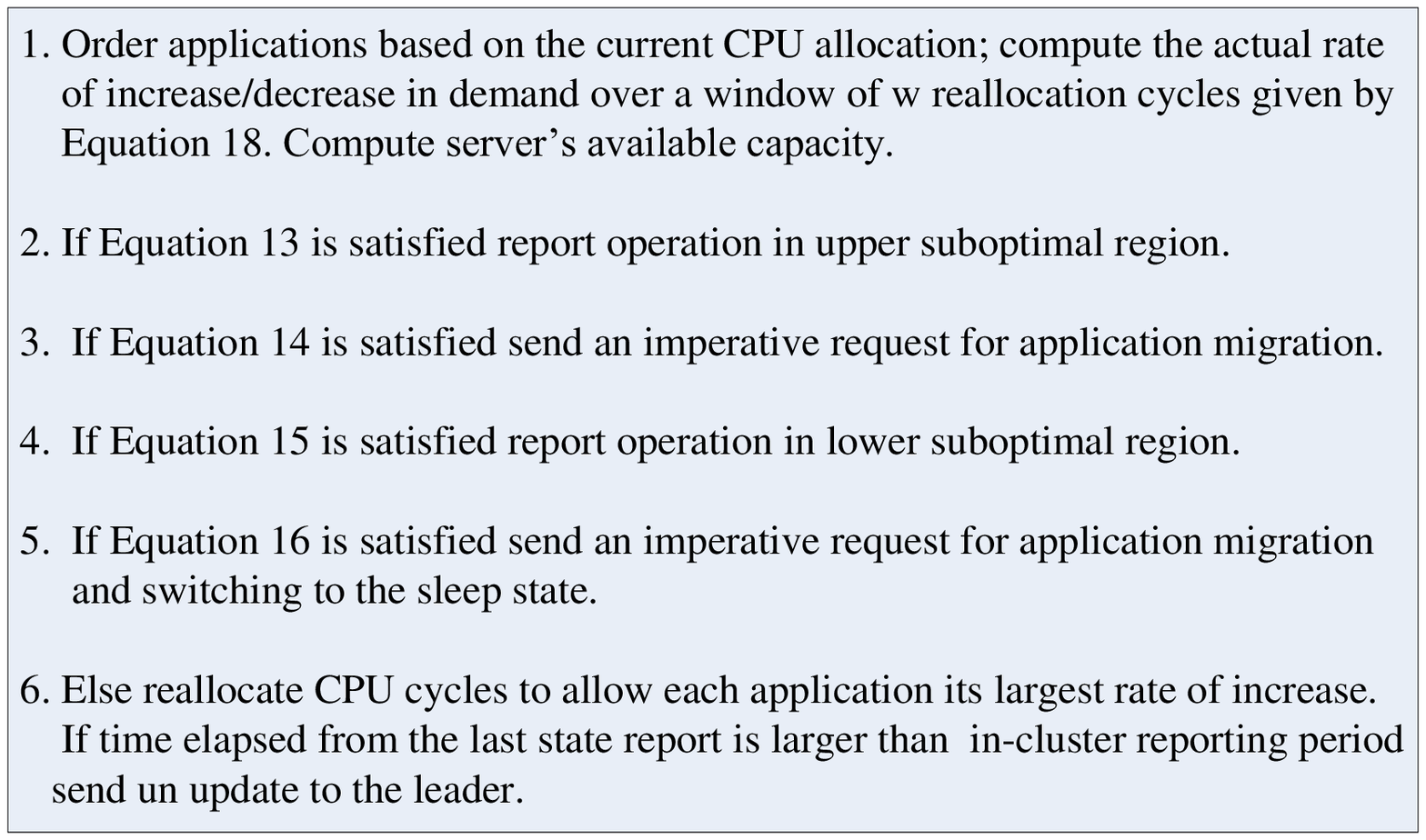}
\end{center}
\caption{Synchronous reallocation algorithm used by the SAM running on server $\mathcal{S}_{k}$.}
\label{ServerAlgorithm}
\end{figure*}

\medskip

\underline{Server Application Management algorithms.}
Server $\mathcal{S}_{k}$ calculates the additional demand for CPU cycles
of individual applications over the last reallocation cycle

\begin{equation}
c_{i,k}(t) = a_{i,k}(t) - a_{i,k} (t - \tau_{k})
\end{equation}
and over a window of $w$ reallocation intervals

\begin{equation}
c_{i,k}^{w}(t) = \sum_{j=0}^{w-1} c_{i,k} (t -j\tau_{k}).
\end{equation}

The additional demand for CPU cycles for all applications running on server $\mathcal{S}_{k}$
over the last reallocation cycle is

\begin{equation}
c_{k}(t) = \sum_{i} c_{i,k}(t)
\end{equation}
and over a window of $w$ reallocation intervals

\begin{equation}
c_{k}^{w}(t) = \sum_{j=0}^{w-1} c_{k} (t -j\tau_{k}).
\end{equation}

Server $\mathcal{S}_{k}$ maintains a control data structure including all currently
running application ordered by CPU cycle consumption. The
application record of $\mathcal{A}_{i}$ includes:

\smallskip

$a_{i}(t)$ -  current demand for CPU cycles.

\smallskip

$c_{i,k}(t)$ - change in demand over last reallocation cycle.

\smallskip

$c_{i,k}^{w}(t)$ - change in demand over window $w$.

\smallskip

$\lambda_{i}$ - highest rate of increase in demand.

\smallskip

$p_{i,k}(t)$ - cost of migration.

\smallskip

$q_{i,k}(t)$  - cost of horizontal scaling.

\smallskip

The SAM component of the VMM running on server $\mathcal{S}_{k}$ operates synchronously and asynchronously, in response to interactions with the leader, $\mathcal{L}_{\mathcal{C}}$ and other servers accepting the migration of an application currently running on $\mathcal{S}_{k}$.

A. {\it The reallocation algorithm executed every $\tau_{k}$ units of time}, see
Figure \ref{ServerAlgorithm}.

\smallskip

A report of operation in the lower suboptimal region consists of the list
of all applications running on $\mathcal{S}_{k}$. Similarly, a report of operation 
in the upper suboptimal region consists of a list
of $\nu$  applications, server $\mathcal{S}_{k}$ recommends to be migrated. The
applications included in this list satisfy two conditions: (i) ensure that after
migration the server $\mathcal{S}_{k}$ will operate in the optimal region, $\mathcal{O}_{k}$,
and (ii) the cost of migration of the $\nu$ applications

\begin{equation}
\sum_{i=1}^{\nu} \left( p_{i,k}(t) + q_{i,k}(t) \right)
\end{equation}
is minimum. The information supplied for each one of the application is the same
as in case of operation in the lower suboptimal region.

\medskip

B. {\it Asynchronous operation.}

\smallskip

B.1 When $\mathcal{S}_{k}$ receives a request from $\mathcal{L}_{\mathcal{C}}$ to accept the
migration or vertical scaling on an application, it first checks that by accepting
the request it will still be operating on an optimal region. If so, it sends an accept message
to the leader and to $\mathcal{S}_{v}$, the server requesting the migration or vertical scaling of application. In the former case it starts one or more VMs for the application;
in the latter case it waits to receive from $\mathcal{S}_{v}$  the snapshot of the VM image and then starts the new VMs.

\smallskip

B.2 When, in response to a report of operation in a suboptimal region, server $\mathcal{S}_{k}$ receives an accept message for vertical scaling of application $\mathcal{A}_{i,k}$ from
another server, $\mathcal{S}_{v}$, it stops the application, constructs the image of the VM
running the application and then sends it to $\mathcal{S}_{v}$. For horizontal scaling, it sends
 $\mathcal{S}_{v}$ the location of the image.

\medskip

\underline{Cluster Leader algorithms.}
The leader performs two basic functions: admission control for new applications and the management of servers in the cluster. For the later function the leader acts as a broker, once it receives a request for in-cluster scaling it identifies the potential target and then the two servers, the one sending the request, and the one accepting to be the target for horizonal or vertical scaling; once an agreement has been reached the two servers carry out the operation without the intervention of the leader.

Once $\mathcal{L}_{\mathcal{C}}$ receives from $\mathcal{S}_{k}$ either a number of $r^{l}$,  warning of operation in the lower suboptimal region, or an imperative request, the applications running on the server are migrated to other servers and the server is switched to the standby/sleep state. Similarly, after $r^{h}$ warning of operation in the upper suboptimal region, or an imperative request, the applications designated by $\mathcal{S}_{k}$ are migrated to other servers. The choice of $r^{l}$ and $r^{h}$ allows servers to operate in a suboptimal region for brief periods of time thus,
reduce the network traffic and the overhead of application migration.

The leader $\mathcal{L}_{\mathcal{C}}$ maintains two control structures:
\begin{enumerate}
\item
{\it SleepingCS} - servers in sleep state; ordered in the increasing order of computing power reflected by the constant $\gamma_{k}$.
\item
{\it RunningCS} - running servers; ordered in the increasing order of computing power. Within a group of servers with similar $\gamma_{k}$, the servers are ordered in the increasing order of available capacity.
\end{enumerate}

{\it Admission control.} When $\mathcal{L}_{\mathcal{C}}$ receives a request to accept
a new application $\mathcal{A}$ it computes the available capacity
\begin{equation}
d_{\mathcal{C}}(t) = \sum_{k=1}^{n_{\mathcal{C}}} d_{k}(t)
\end{equation}
and admits $\mathcal{A}$ if the system is not overloaded:

\begin{equation}
{ {d_{\mathcal{C}}(t)} \over {\sum_{k=1}^{n_{\mathcal{C}}} \gamma_{k}}} \le 0.8
\end{equation}

{\it Selection of a target for horizontal or vertical scaling of application $\mathcal{A}$.}
The first step is to classify the application based on the evolution of its CPU cycle needs over the past window of
$w$ reallocation cycles in several categories: rapidly increasing resource demands (RI), moderately increasing (MI), stationary (S), moderately decreasing (MD), and rapidly decreasing (RD). The target selection is guided by two objectives:

\smallskip

(i) Ensure that the target server selected will be able to accommodate application scaling for an extended period of time, while operating in its optimal region; this will help reduce the migration costs and the power consumption.

\smallskip

(ii) Keep the user costs low by selecting the least costly server, the server with the lowest $\gamma_{k}$ that satisfies condition (i).

\medskip

The strategies for the five classes of applications are:

\noindent \underline{\it RI, MI} - Consider a window of $\phi_{RI}$ and respectively $\phi_{MI} << \phi_{RI}$ future intervals and determine the largest possible increase in resource demand of application $\mathcal{A}_{i}$. For the {\it RI} case

\begin{equation}
a_{i}(t+ \phi_{RI} \times \lambda_{i}) = a_{i}(t) + \phi_{RI} \lambda_{i}
\end{equation}
Search the {\it RunningCS} list to identify a server $\mathcal{S}_{v}$ with suitable available capacity

\begin{equation}
d_{v}(t) >  a_{i}(t+ \phi_{RI} \times \lambda_{i}).
\end{equation}
If such a server does not exist then wake up  server $\mathcal{S}_{u}$ from {\it SleepingCS} list with the lowest
$\gamma_{u}$; select it as a target if it satisfies the conditions

\begin{equation}
\begin{array} {c}
a_{i}(t) \ge \gamma_{u} \alpha_{u}^{sopt,low}\\
\text{and} \\
a_{i}(t+ \phi_{RI} \times \lambda_{i}) \le \alpha_{u}^{sopt,low}.
\end{array}
\end{equation}
Else continue with the next server on the {\it SleepingCS} list. For the {\it MI} case
use the same procedure with $\phi_{MI}$ instead of  $\phi_{RI}$ and $\lambda_{i}/2$ instead of $\lambda_{i}$.

\medskip

\noindent \underline{\it S} - Search the {\it RunningCS} list to identify a server $\mathcal{S}_{v}$ such that
\begin{equation}
d_{v}(t) >  a_{i}(t).
\end{equation}
If such a server does not exist then wake up  server $\mathcal{S}_{u}$ from {\it SleepingCS} list with the lowest
$\gamma_{u}$; select it as a target if it satisfies the conditions

\begin{equation}
\begin{array} {l}
a_{i} \ge \gamma_{u} \alpha_{u}^{sopt,low}.
\end{array}
\end{equation}

\noindent \underline{\it MD, RD} -  Consider a window of $\phi_{RD}$ and, respectively $\phi_{MD} << \phi_{RD}$ future intervals and determine the largest possible decrease in resource demand of application $\mathcal{A}_{i}$. In the {\it RI} case

\begin{equation}
a_{i}(t+ \phi_{RD} \times \lambda_{i}) = a_{i}(t) - \phi_{RD} \lambda_{i}.
\end{equation}
To identify a server $\mathcal{S}_{v}$ with suitable available capacity $d_{v}(t)$ search the {\it RunningCS} list, where

\begin{equation}
d_{v}(t) >  a_{i}(t+ \phi_{RD} \times \lambda_{i}).
\end{equation}
If such a server does not exist then wake up  server $\mathcal{S}_{u}$ from {\it SleepingCS} list with the lowest $\gamma_{u}$; select it as a target if it satisfies the conditions

\begin{equation}
\begin{array} {c}
a_{i}(t) \ge \gamma_{u} \alpha_{u}^{sopt,low}\\
\text{and} \\
a_{i}(t+ \phi_{RI} \times \lambda_{i}) \le \alpha_{u}^{sopt,low}.
\end{array}
\end{equation}
Else continue with the next server on the {\it SleepingCS} list. For the {\it MI} case
use the same procedure with $\phi_{MD}$ instead of  $\phi_{RD}$ and $\lambda_{i}/2$ instead of $\lambda_{i}$.

\medskip

The algorithms described in this section assume that the thresholds for normalized performance and power consumption of server $\mathcal{S}_{k}$ are constants. If the processor supports dynamic voltage and frequency scaling thus, it is capable to increase or decrease the operating voltage or frequency of a processor to increase the instruction execution rate and, respectively, to decrease it and conserve power, these thresholds, $\alpha_{k}^{opt,low}(t), \alpha_{k}^{opt, high}(t), \alpha_{k}^{sopt,low}(t), \alpha_{k}^{sopt, high}(t)$, $\beta_{k}^{opt,low}(t), \beta_{k}^{opt, high}(t), \beta_{k}^{sopt,low}(t), \beta_{k}^{sopt, high}(t)$, will vary in time.

The basic philosophy will be the same, we shall attempt to
keep every server in an optimal operating region. An additional complication of the algorithms is that we have to determine if it is beneficial to increase/decrease the power used thus, push up/down the thresholds of the operating regions of the server. We still want to make most scaling decisions locally.

When the demand for CPU cycles increases, the SAM
must compare the additional power consumption of $\mathcal{S}_{k}$ with the power consumption for migration and the power
consumption at an average power consumption of all servers in $\mathcal{C}$. The leader $\mathcal{L}_{\mathcal{C}}$ should then attempt to identify a target server $\mathcal{S}_{v}$ operating below this average level and direct migration of
the application to that server. When the demand decreases, the SAM must decide if by reducing the voltage
and/or frequency the server will still be able to operate in an optimal region with the lower load.

\section{A Simulation Experiment}
\label{SimulationExperiment}

To evaluate the algorithms discussed in Section \ref{ScalingAlgorithms} we conduct a simulation study; the study will give us some indications about the operation of the algorithm in clusters of different sizes and of the effectiveness and of the overhead of the algorithm. Simulation studies are also important for determining optimal choices for various parameters of the algorithms such as $\tau_{k}, \tau_{i}, r^{l}, r^{h}$ and $w$.

\begin{figure*}[!ht]
\begin{center}
\includegraphics[width=7.5cm]{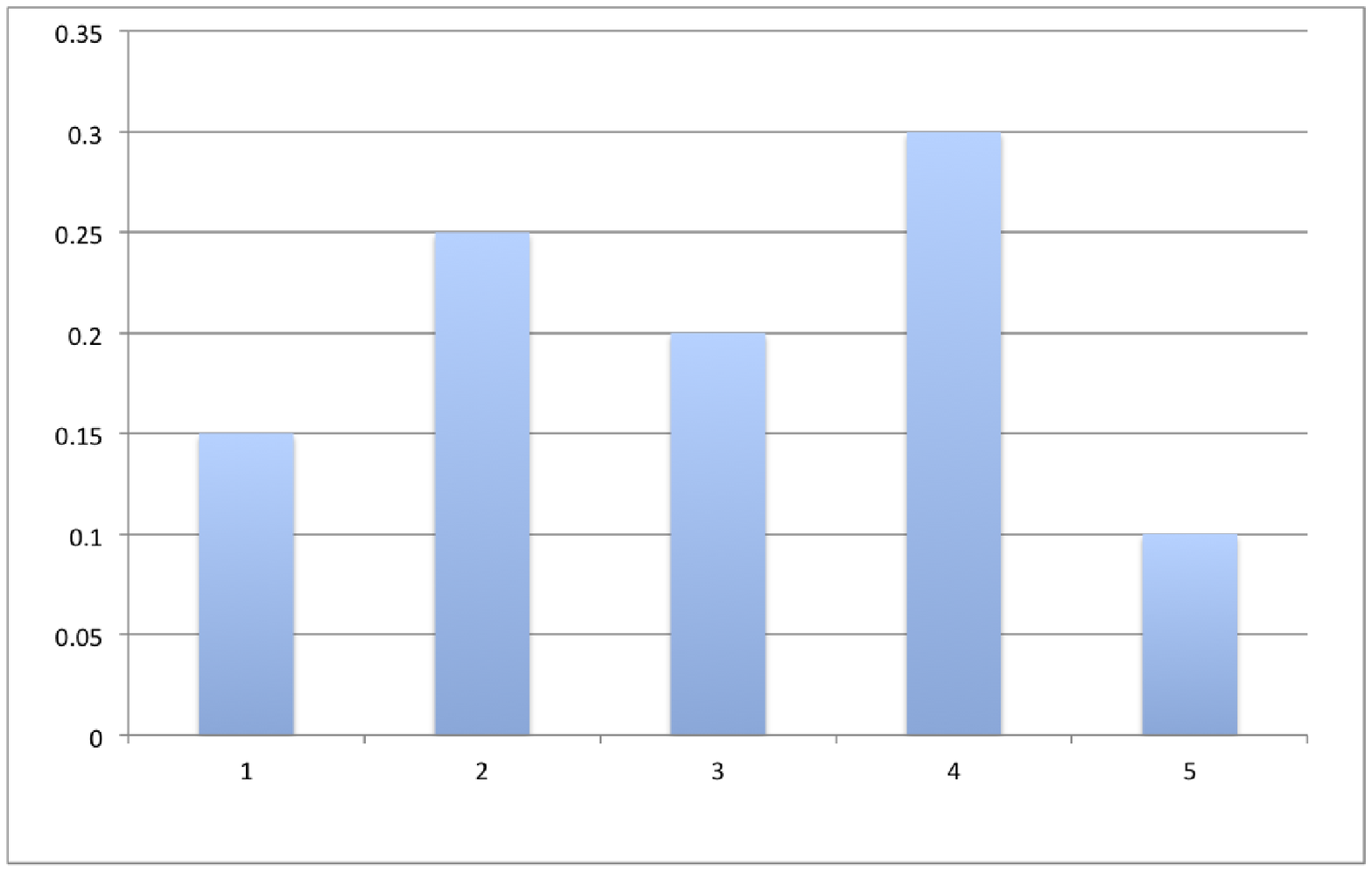}
\includegraphics[width=7.5cm]{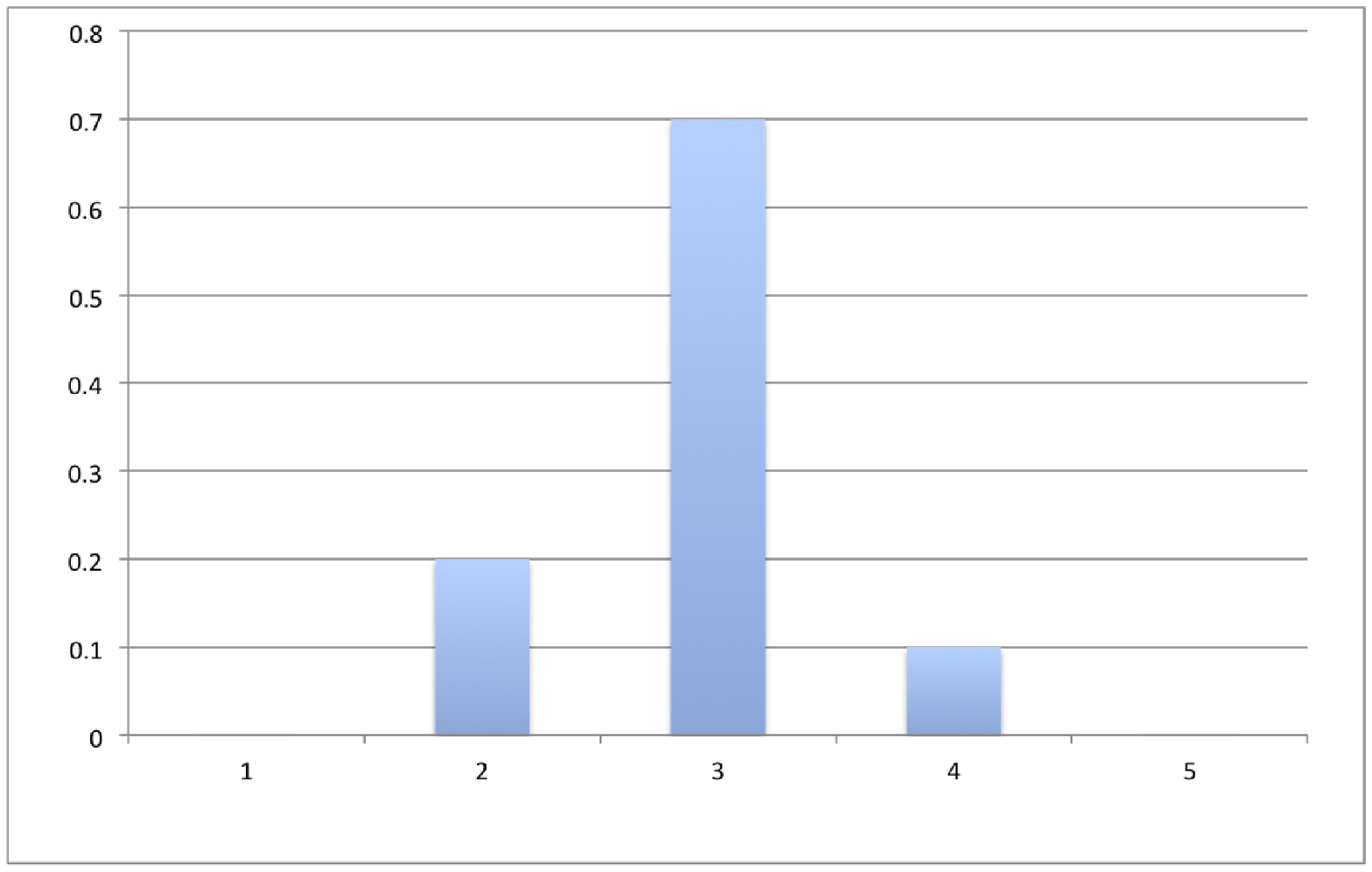}\\
~~~Initial~~~~~~~~~~~~~~~~~~~~~~~~~~~~~~~~~~~~Typical~~~~~~\\
\includegraphics[width=10cm]{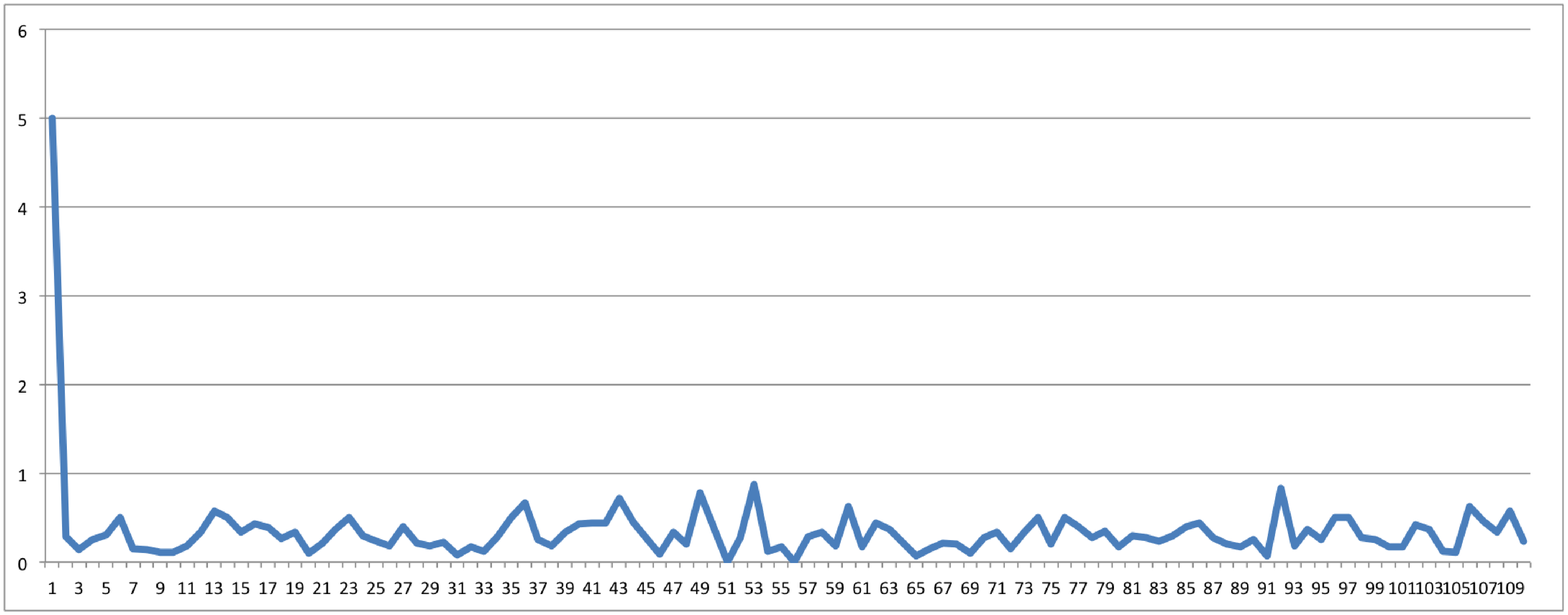}
\end{center}
\caption{Small cluster with 20 servers; average cluster load $50\%$. (Top) Initial and typical distribution of the number of servers in the five operating regions. (Bottom) The ratio of in-cluster to local decisions in response to scaling requests versus time.}
\label{ClusterSize20}
\end{figure*}

The metrics for assessing the effectiveness and the overhead of the algorithms are:

\begin{enumerate}
\item
The evolution of the number of servers in each of the five operating regions as a result of the load migration mandated by the algorithm; from the point of view of power consumption  the five regions are: optimal, suboptimal low/high, and undesirable low/high. This evolution measures the effectiveness of the algorithm.
\item
The average number of servers in each spent by the servers in each of the 
\item
The ratio of local versus in-cluster scaling decisions during simulation. This reflects the overhead of the algorithm.
\end{enumerate}

\begin{figure*}[!ht]
\begin{center}
\includegraphics[width=7.5cm]{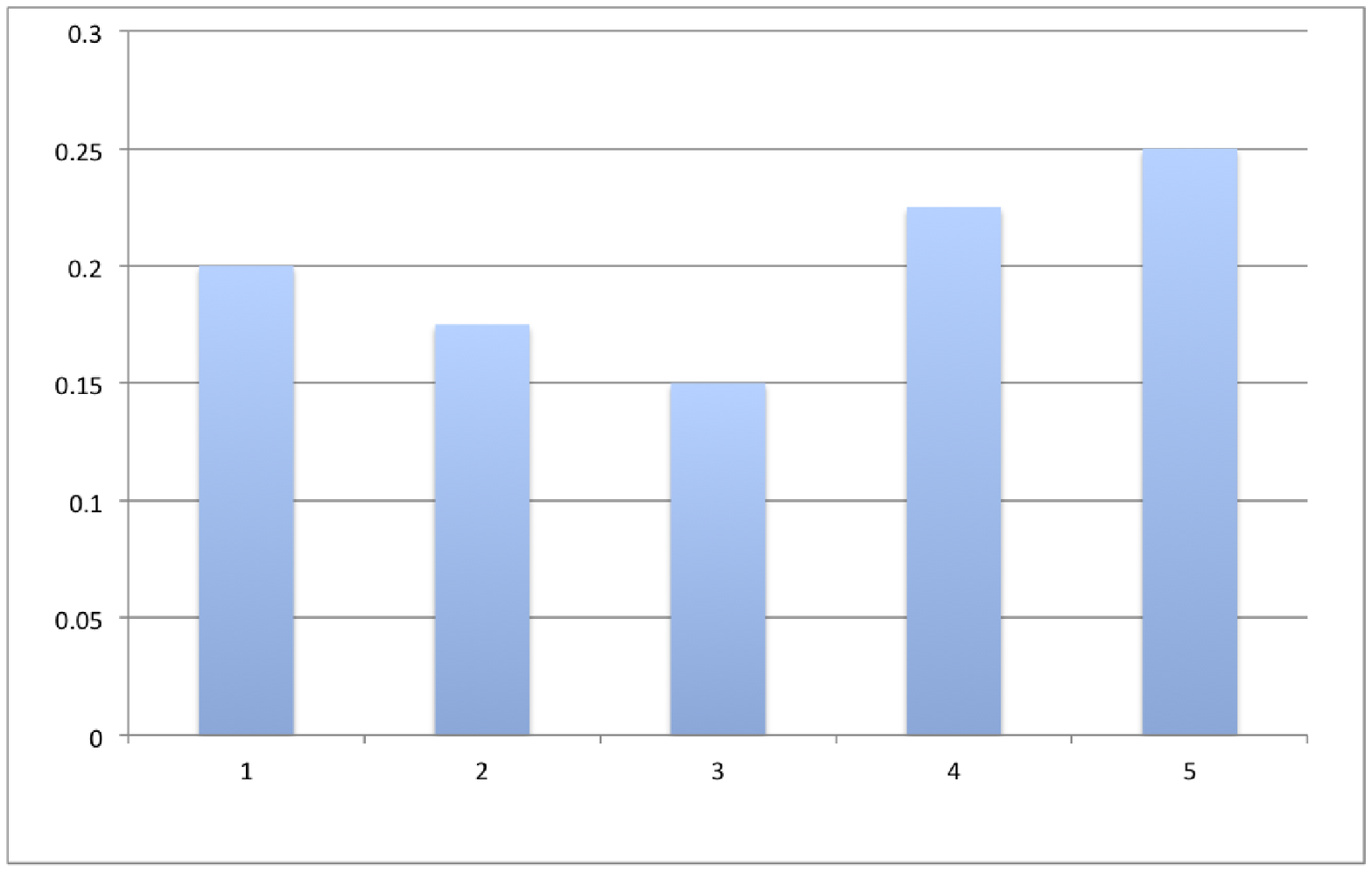}
\includegraphics[width=7.5cm]{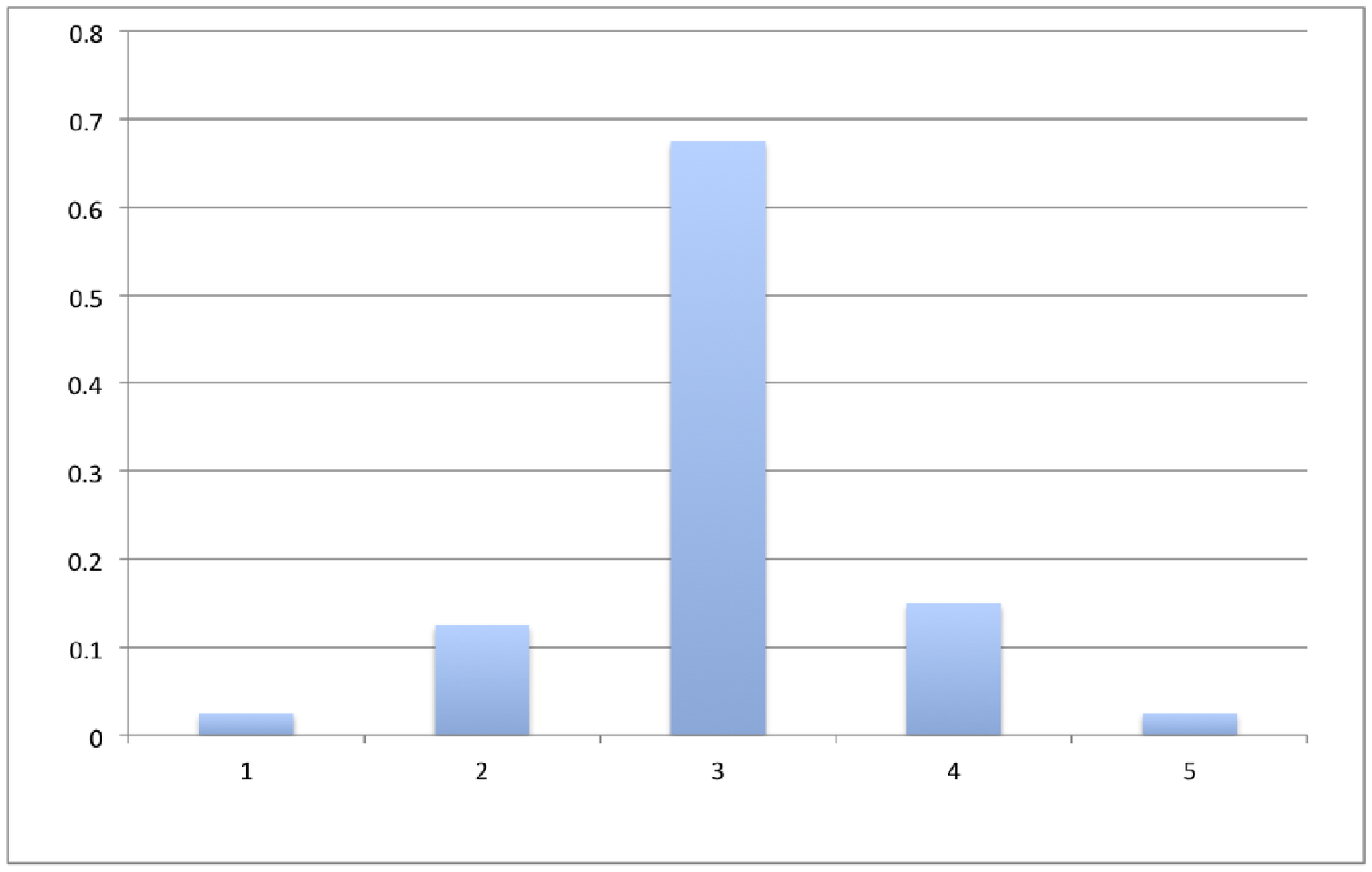}\\
~~~Initial~~~~~~~~~~~~~~~~~~~~~~~~~~~~~~~~~~~~Typical~~~~~~\\
\includegraphics[width=16cm]{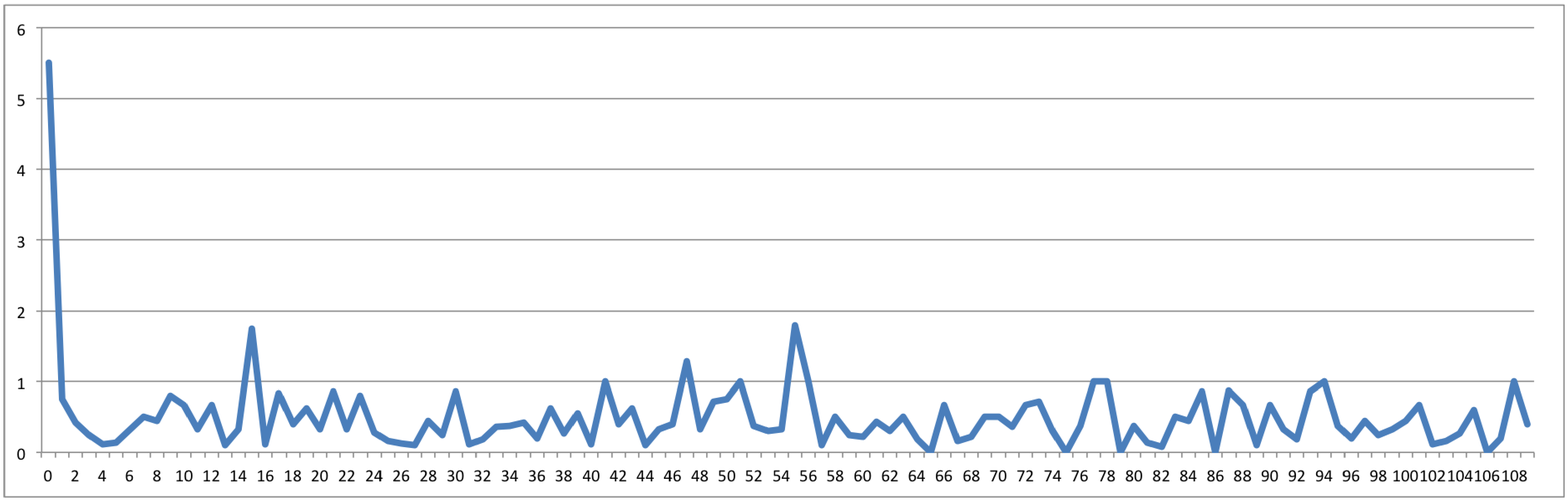}
\end{center}
\caption{Small to medium size cluster with 40 servers; the average cluster load is $50\%$. (Top) Initial and typical distribution of the number of servers in the five operating regions. (Bottom) The ratio of in-cluster to local decisions in response to scaling requests versus time.}
\label{ClusterSize40}
\end{figure*}

\begin{figure*}[!ht]
\begin{center}
\includegraphics[width=17cm]{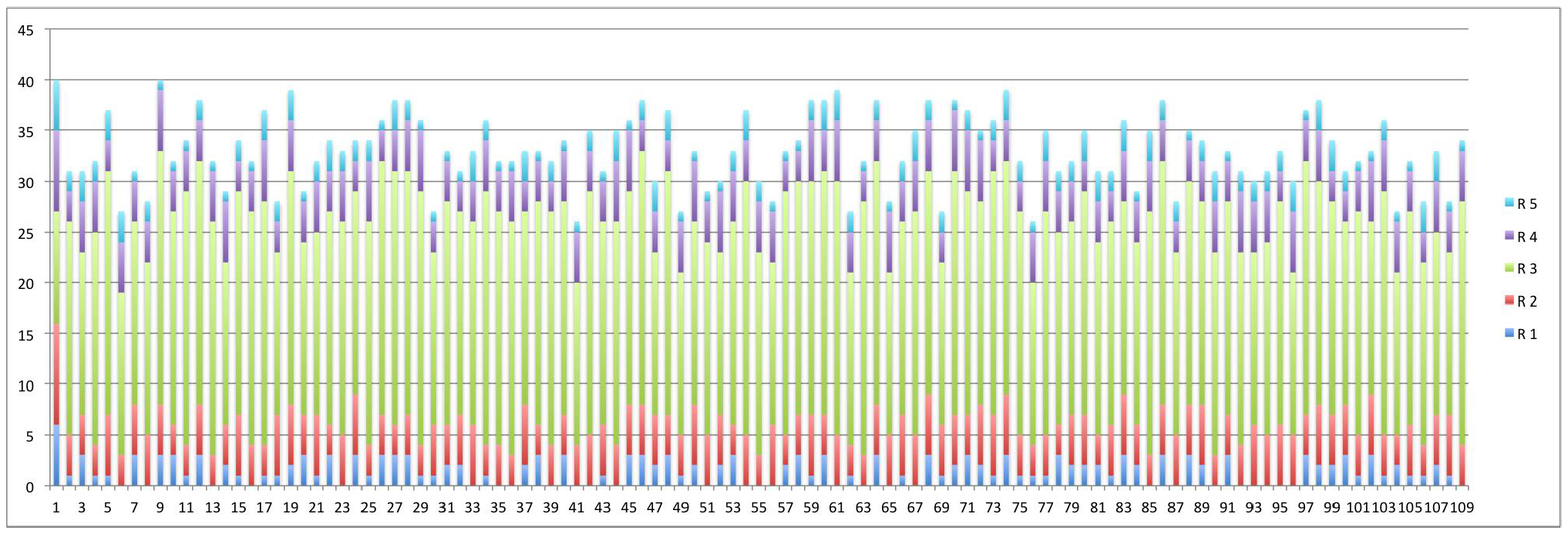}
\end{center}
\caption{The number of servers in each of the five operating regions during the entire  simulation. The cluster has 40 servers and its average load is $50\%$.}
\label{ServersVsRegion}
\end{figure*}

In our simulation experiments we have considered clusters of size $20, 40, 60, 80$ and $100$ servers and then we have assigned a random number of applications to each server. $\alpha_{k}^{sopt,low}, \alpha_{k}^{opt,low},\alpha_{k}^{opt,high}$ and $\alpha_{k}^{sopt,high}$, the limits for the five operating regions of each server, $\mathcal{S}_{k}$, are random numbers uniformly distributed in the following intervals: $[0.2 - 0.25], [0.25 - 0.45], [0.55 - 0.8]$ and $[0.8 - 0.85]$, respectively. The requests for scaling are uniformly distributed in the range $3 - 8\%$ of the demand of each application
running on a server

The attributes of each application $\mathcal{A}_{i,k}$, such as $\lambda_{i,k}, p_{i,k}, q_{i,k}$ are also randomly generated.

For the first set of experiments, the initial workload is uniformly distributed in a very broad range, from $20\%$ to $80\%$ of the servers maximum capacity thus, the average cluster workload is $50\%$ of its capacity; the number of servers in the cluster is in the range  $20$ to $100$. We consider first a cluster size of 20. The two Figures \ref{ClusterSize20} (Top)  show, respectively, the initial and the typical distribution of the number of servers in the five operating regions of a server: undesirable low, $\mathcal{R}_{1}$; suboptimal low, $\mathcal{R}_{2}$; optimal, $\mathcal{R}_{3}$; suboptimal high, $\mathcal{R}_{4}$; and undesirable high, $\mathcal{R}_{5}$. We consider that the typical operation was reached after the system evolved past the half of the simulation interval.

These histograms show that in normal operation $70\%$ of the servers are in the optimal region, $20\%$ and, respectively, $10\%$  are in the suboptimal low and high regions. We conclude that in case of a small cluster size, $n_{\mathcal{C}} =20$, the algorithm performs well, as no servers ended up in the undesirable regions.

Figure \ref{ClusterSize20} (Bottom) shows the ratio of in-cluster to local decisions in response to scaling requests. The simulation covers the first $18 \times 60 = 1048$ seconds and the time units on the horizontal axis is $10$ seconds. A ratio of $5$ means that there are  5 times more requests for scaling involving the cluster leader than local decisions. This is the case of the first interval of 10 seconds of the simulation when only 8 out of 48 decisions were made locally; then this ratio dropped to less than one. This means that after the initial transitory period most decisions were local thus, the communication overhead of the algorithm is not excessive.

The initial spike in decisions involving the cluster leader is not unexpected; indeed, initially $15\%$ and $10\%$  of the clusters are operating in the undesirable low and high regions, respectively; thus, they require immediate application migration. Moreover, $25\%$ and, respectively, $35\%$ of the servers were operating the suboptimal low and high regions. As soon as the  servers were forced out of the undesirable operating regions only those facing large scaling requests required interactions with the cluster leader.

\begin{figure*}[!ht]
\begin{center}
\includegraphics[width=7.5cm]{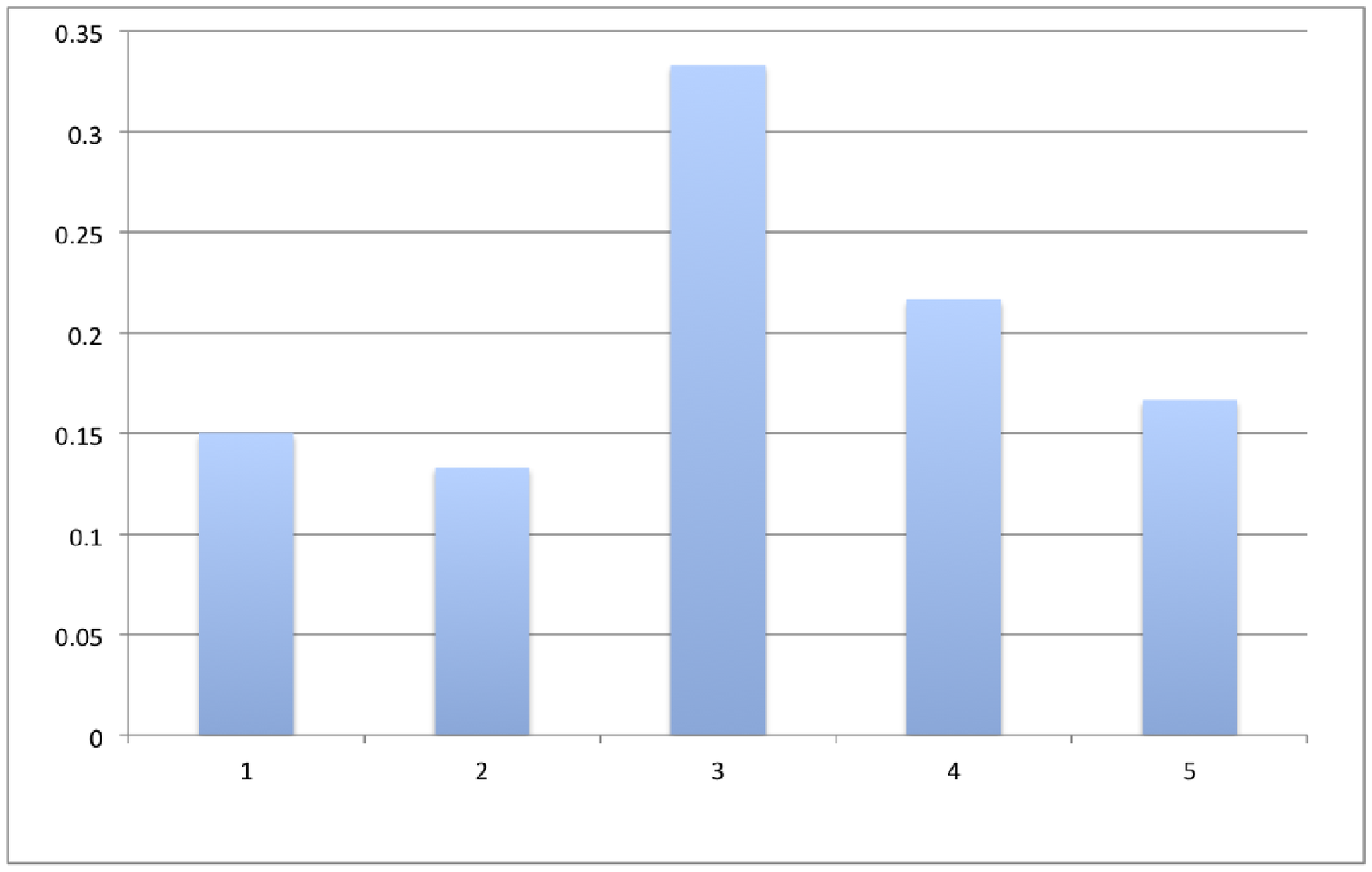}
\includegraphics[width=7.5cm]{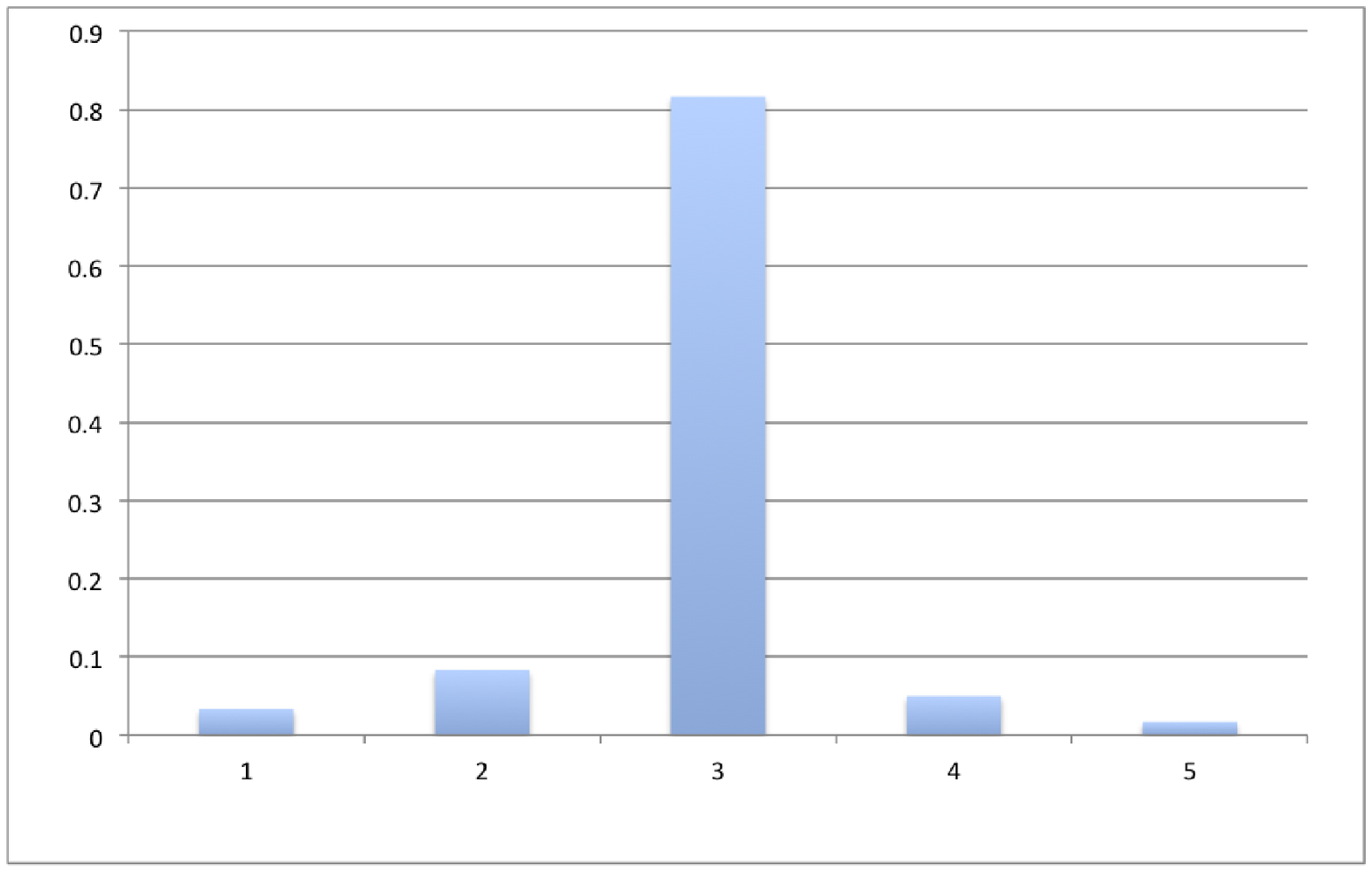}\\
~~~Initial~~~~~~~~~~~~~~~~~~~~~~~~~~~~~~~~~~~~Typical~~~~~~\\
\includegraphics[width=10cm]{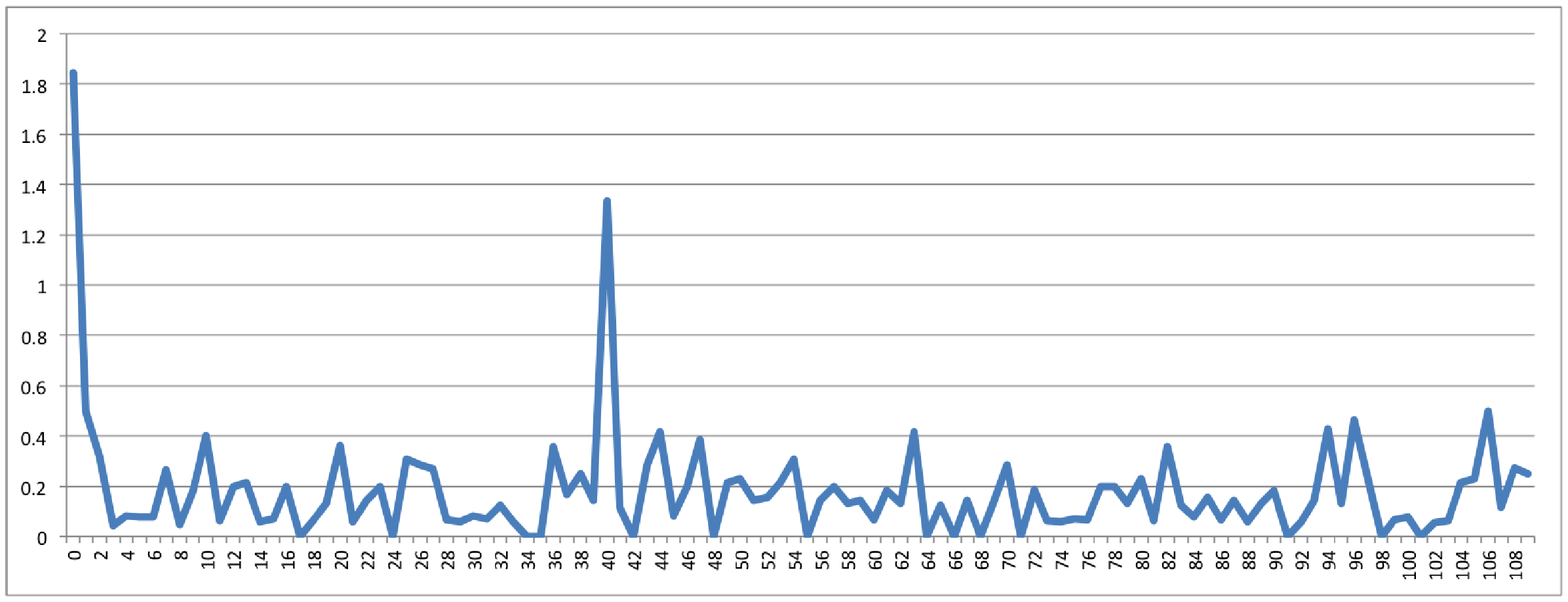}
\end{center}
\caption{Medium size cluster with 60 servers; the average cluster load is $50\%$. (Top) Initial and typical distribution of the number of servers in the five operating regions. (Bottom) The ratio of in-cluster to local decisions in response to scaling requests versus time.}
\label{ClusterSize60}
\end{figure*}

\begin{figure*}[!ht]
\begin{center}
\includegraphics[width=7.5cm]{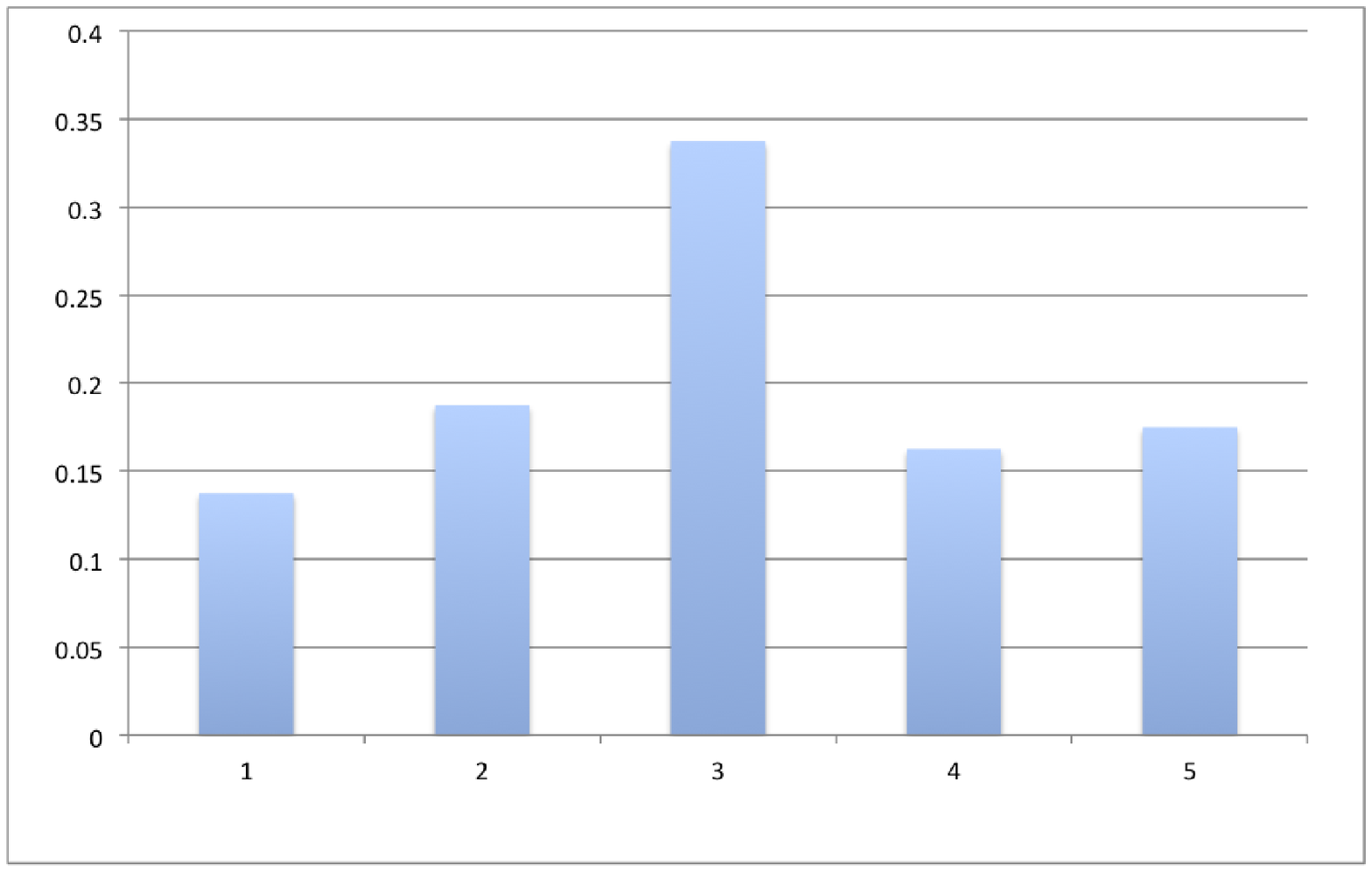}
\includegraphics[width=7.5cm]{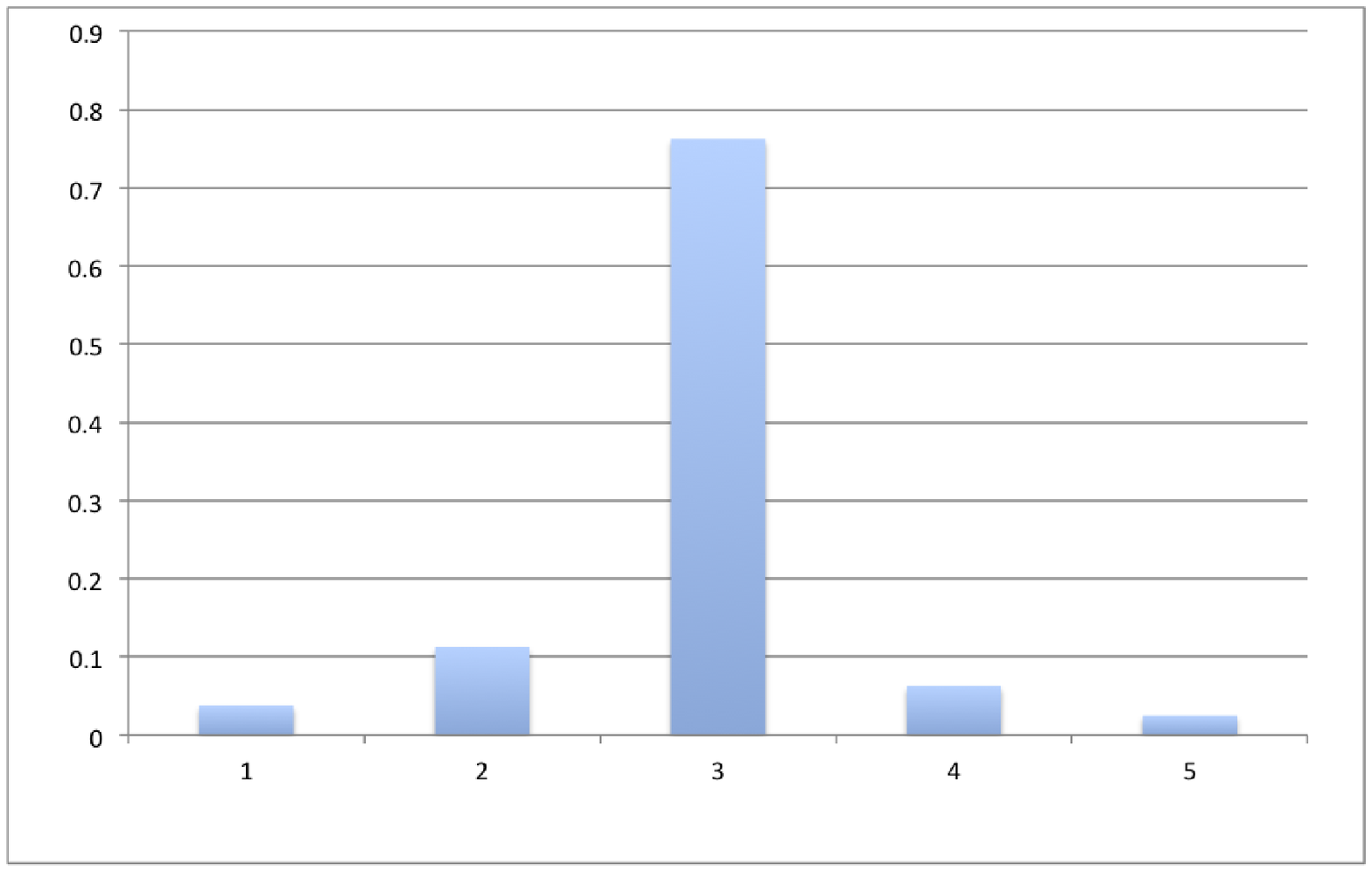}\\
~~~Initial~~~~~~~~~~~~~~~~~~~~~~~~~~~~~~~~~~~~Typical~~~~~~\\
\includegraphics[width=10cm]{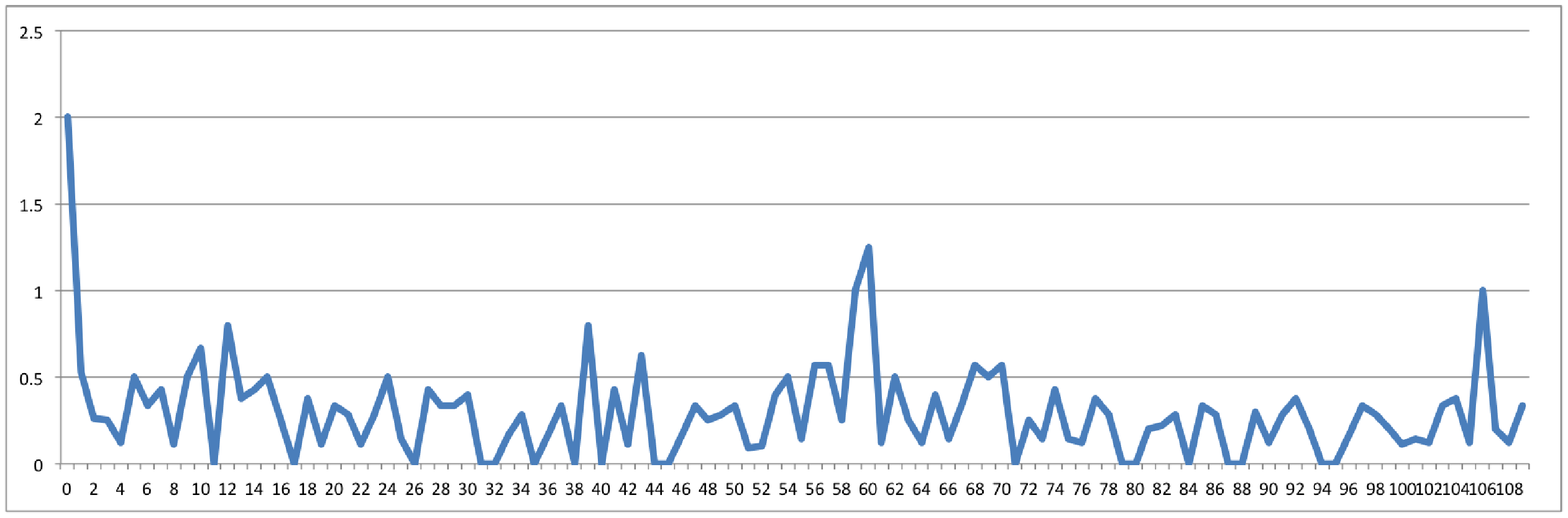}
\end{center}
\caption{Large cluster with 80 servers; the average cluster load is $50\%$. (Top) Initial and typical distribution of the number of servers in the five operating regions. (Bottom) The ratio of in-cluster to local decisions in response to scaling requests versus time.}
\label{ClusterSize80}
\end{figure*}

Figure \ref{ServersVsRegion} shows the number of servers in each of the five operating regions during the simulation when the cluster size is $40$ and the average load placed on the clusters is $50\%$ of their capacity.

\begin{figure*}[!ht]
\begin{center}
\includegraphics[width=7.5cm]{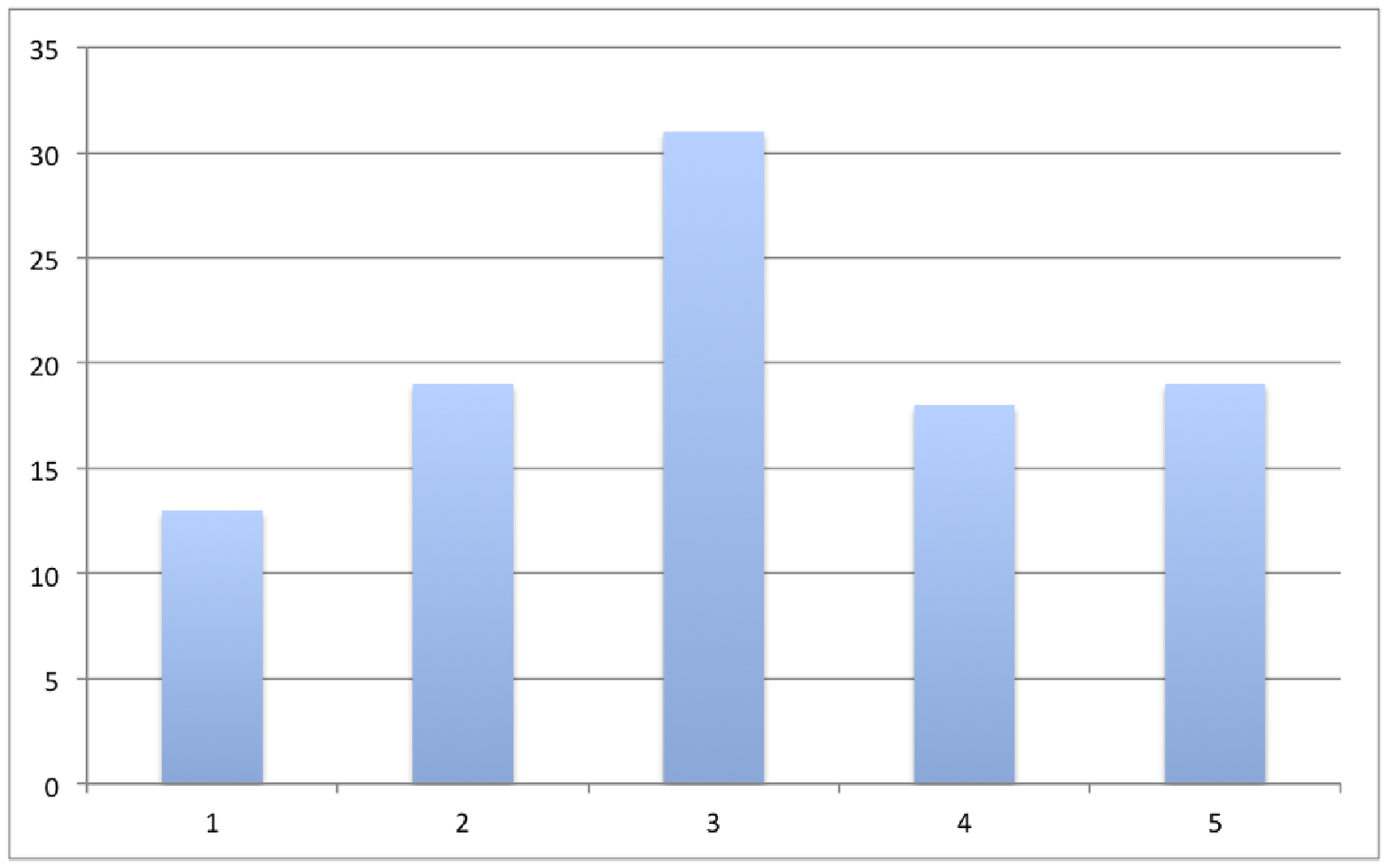}
\includegraphics[width=7.5cm]{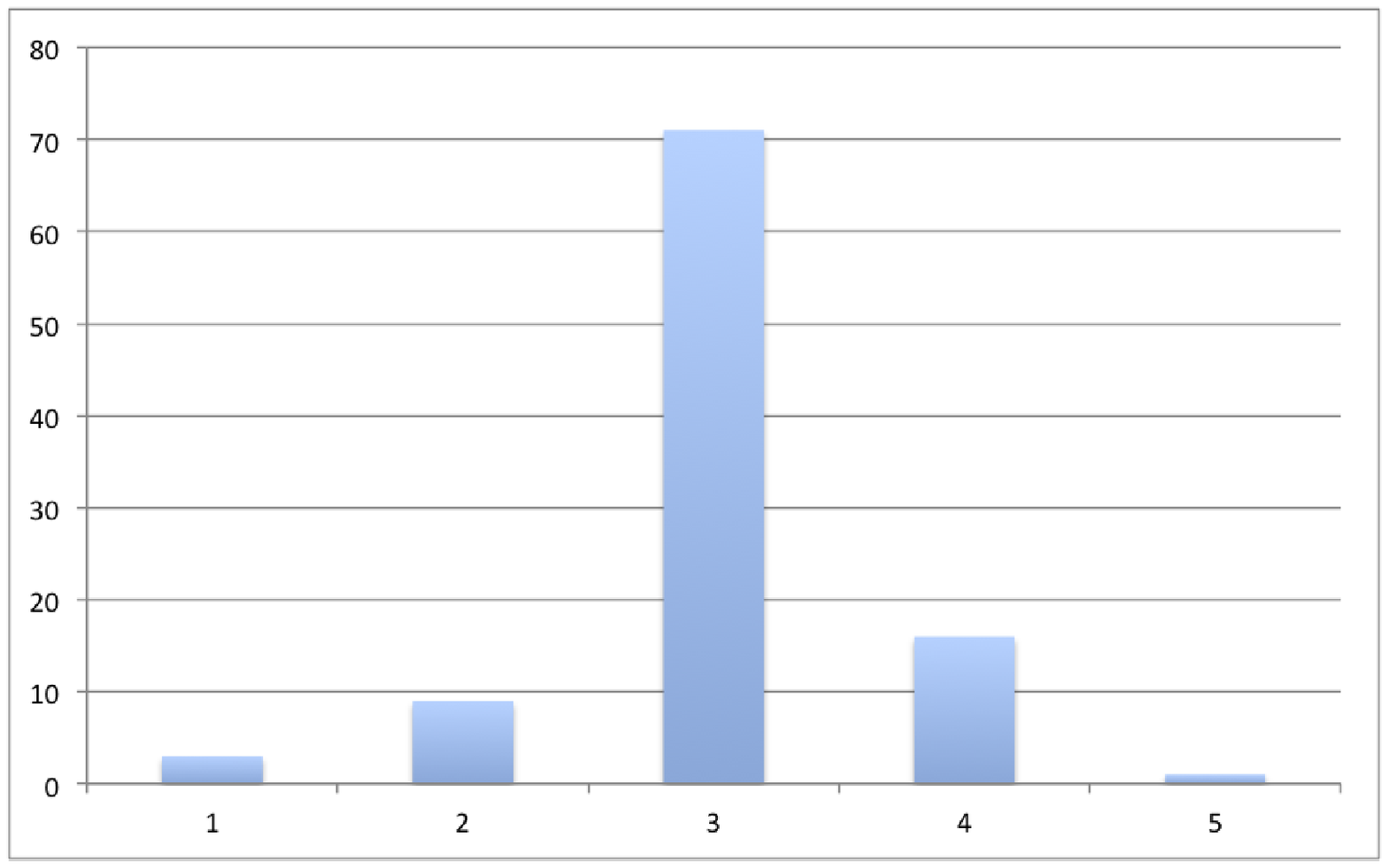}\\
~~~Initial~~~~~~~~~~~~~~~~~~~~~~~~~~~~~~~~~~~~Typical~~~~~\\
\includegraphics[width=10cm]{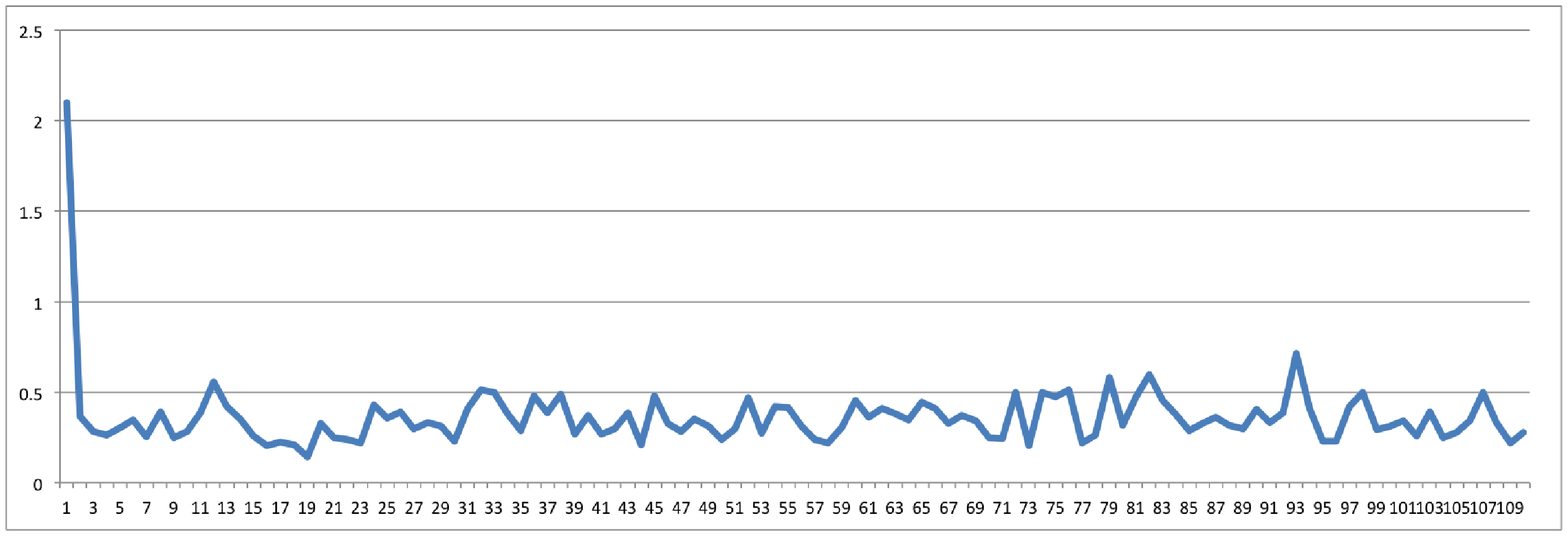}
\end{center}
\caption{Large cluster with 100 servers; the average cluster load is $50\%$. (Top) Initial and typical distribution of the number of servers in the five operating regions. (Bottom) The ratio of in-cluster to local decisions in response to scaling requests versus time.}
\label{ClusterSize100}
\end{figure*}

Figures \ref{ClusterSize40}, \ref{ClusterSize60}, \ref{ClusterSize80}, and  \ref{ClusterSize100} show similar trends  for cluster sizes of $40, 60, 80$ and $100$ servers, respectively. These results allow us to conclude that the algorithm works well for cluster sizes from $20$ to $100$ servers. In all cases during a typical operation less than $5\%$ of the servers operate in one or both undesirable regions, while between $68\%$ and $80\%$ operate in the optimal region.

The algorithm seems to exploit well what we could call a form of {\it locality}; indeed, in a normal operation typically twice as many scaling decisions are made locally, without the involvement of the cluster leader.

The next set of experiments was designed to address the question wether the system load has an effect on the effectiveness and the overhead of the algorithm. We considered two cases: (i) low load -  an initial load uniformly distributed in the interval $20-40\%$ of the server capacity thus, an average load of $30\%$; (ii)  high load - initial server load uniformly distributed in the $60-80\%$ of the server capacity range thus, the average load of the cluster is at $70\%$ of its capacity. We investigated the evolution of the number of servers in each of the five regions and the ratio of in-cluster to local decision for $20, 60$ and $100$ servers in the cluster.

\begin{figure*}[!ht]
\begin{center}
\includegraphics[width=7.5cm]{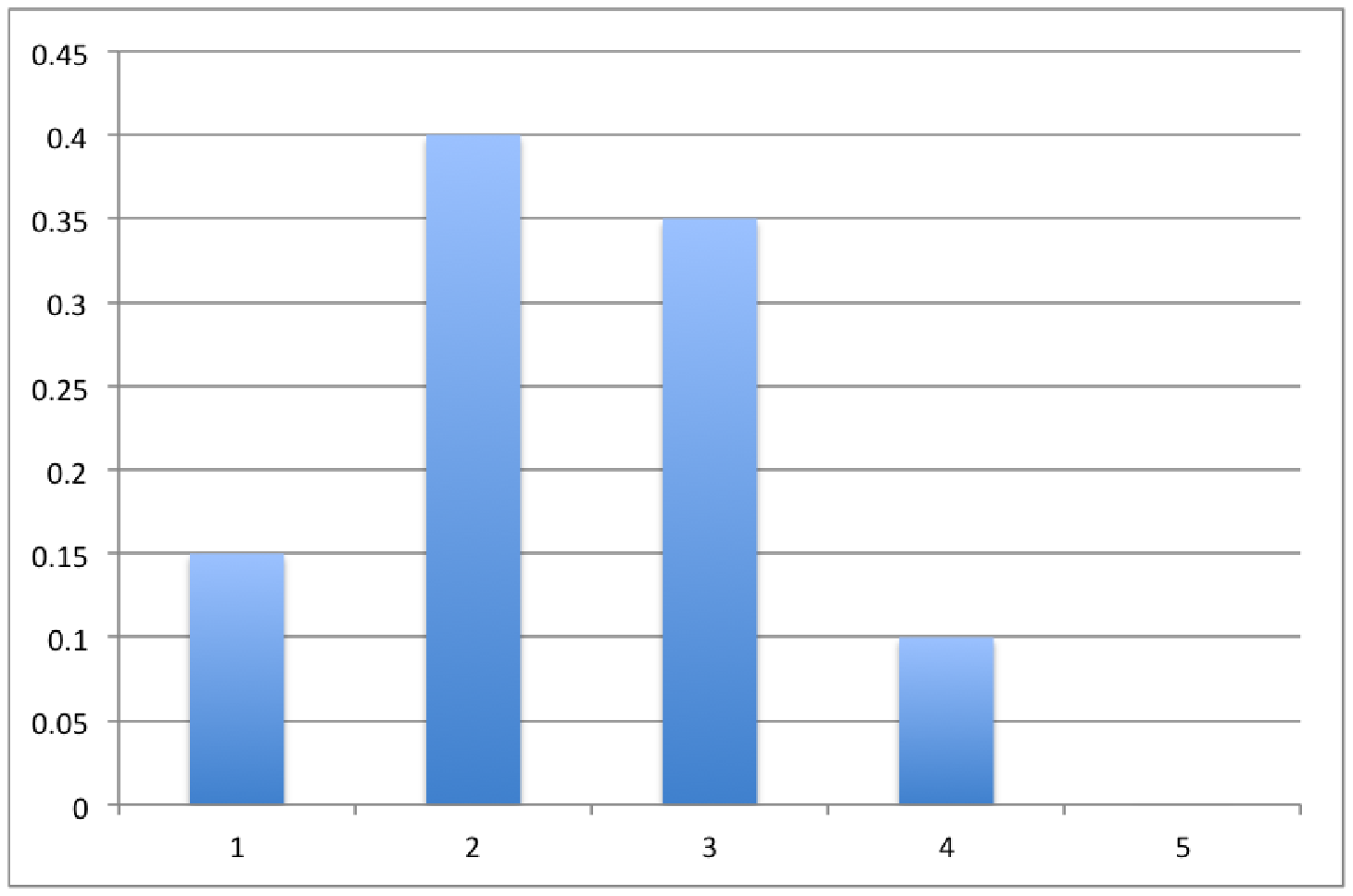}
\includegraphics[width=7.5cm]{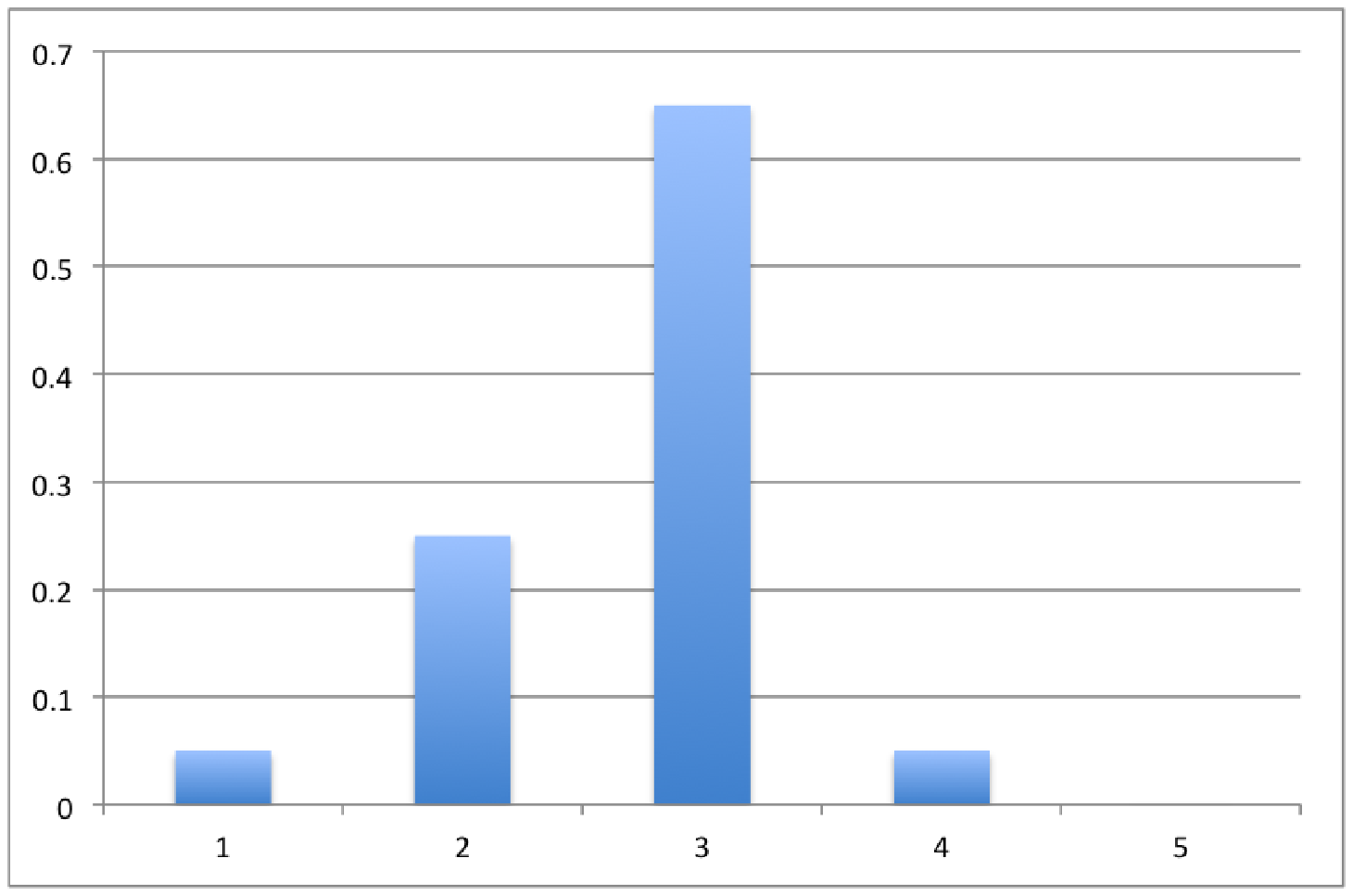}\\
\includegraphics[width=7.5cm]{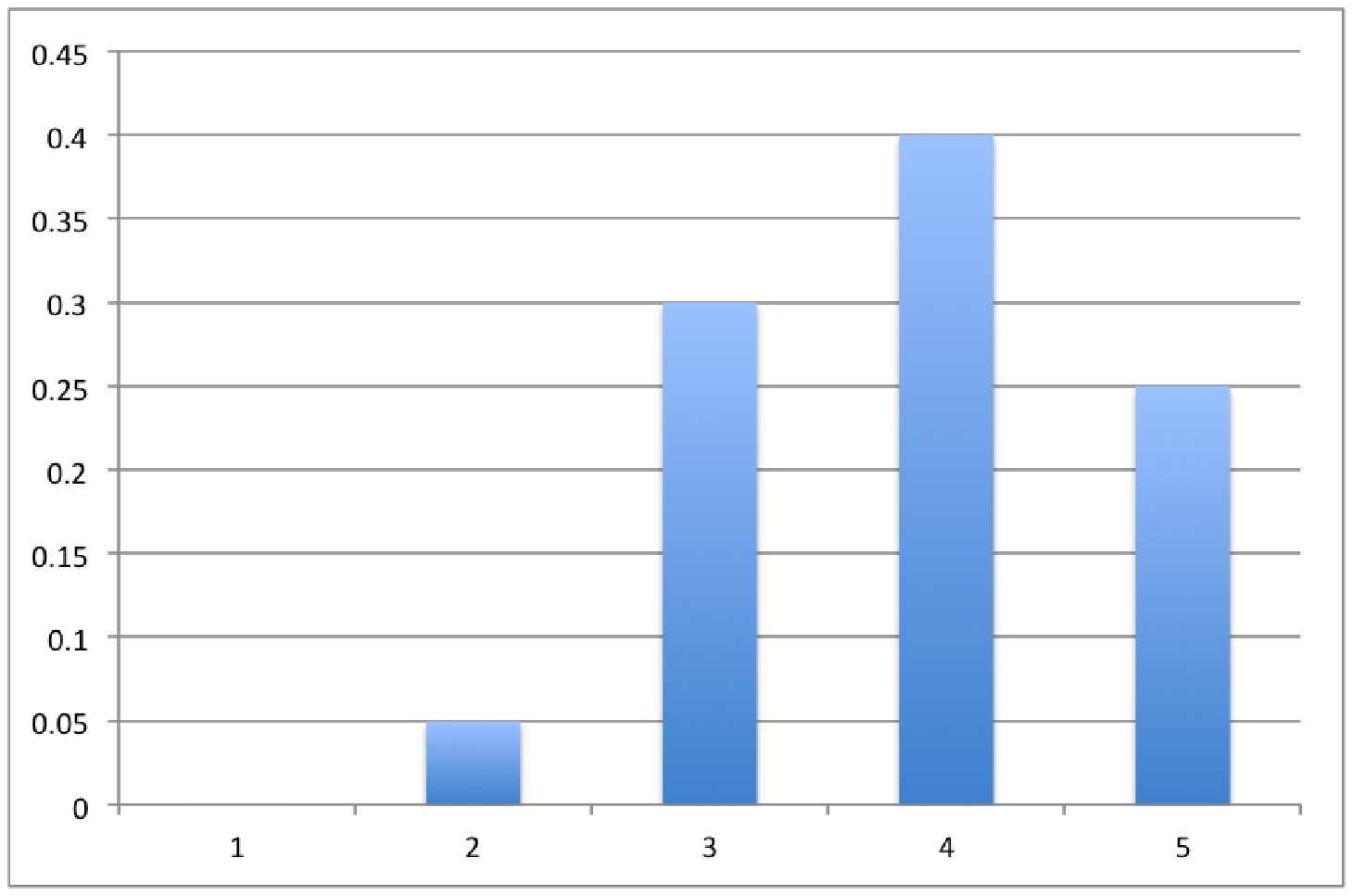}
\includegraphics[width=7.5cm]{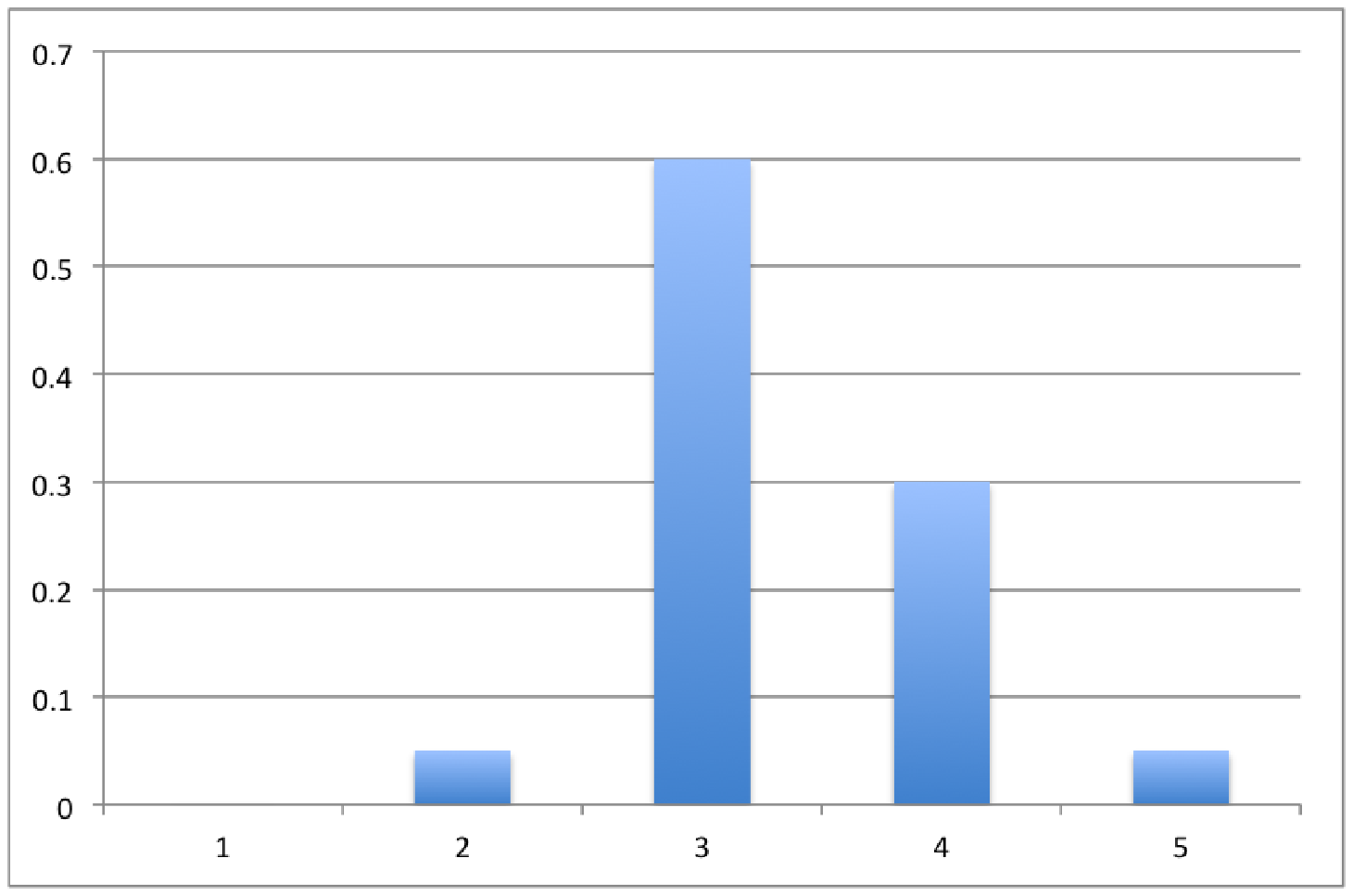}\\
~~~~~~~~~~Initial~~~~~~~~~~~~~~~~~~~~~~~~~~~~~~~~~~~~~~~~~~~Typical
\includegraphics[width=12cm]{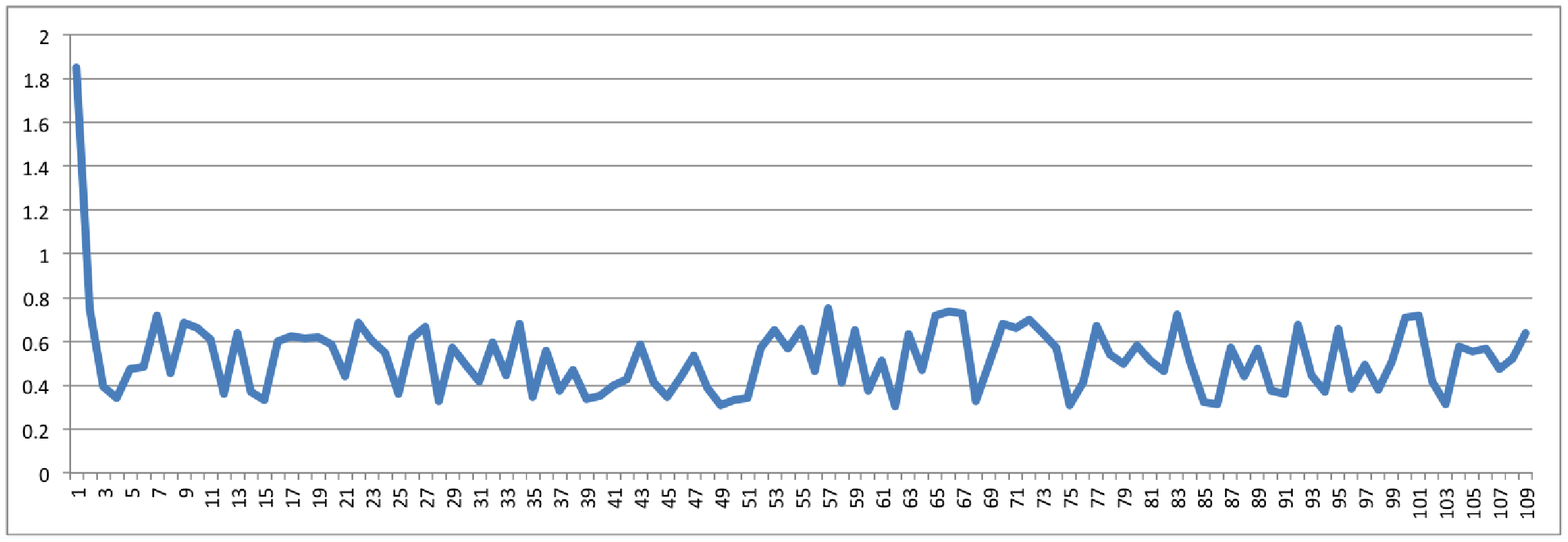}
\includegraphics[width=12cm]{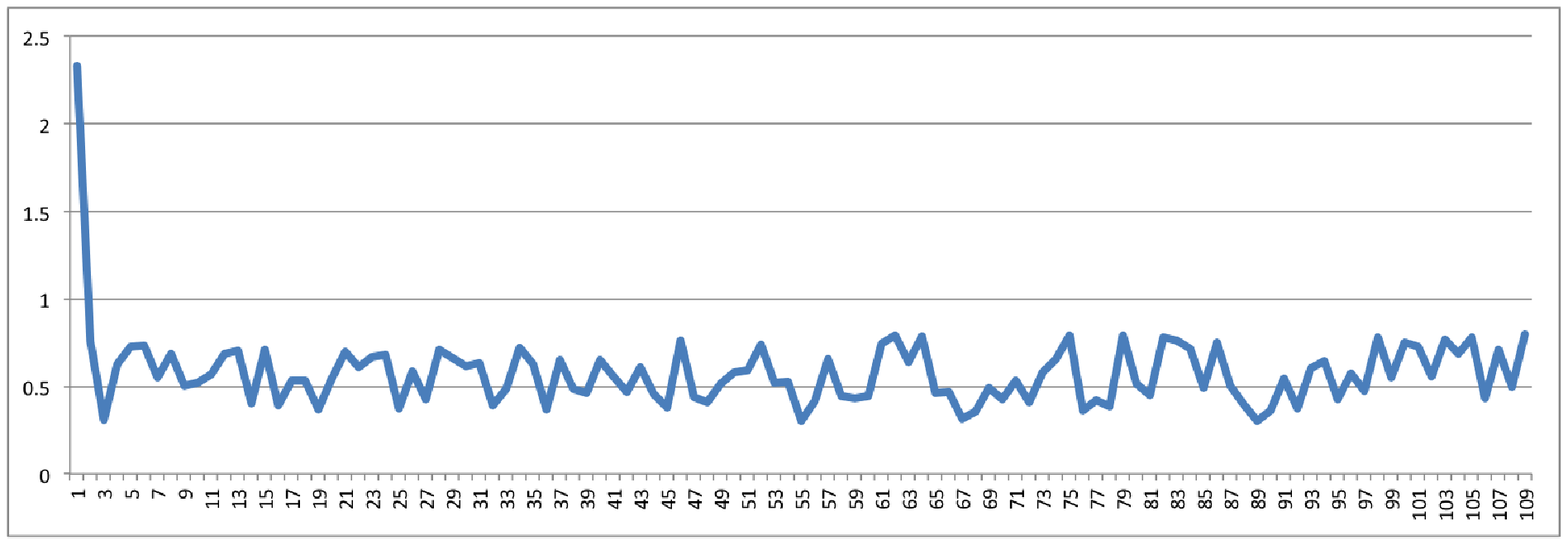}
\end{center}
\caption{The effect of the initial server load when there are 20 servers in the cluster. (a) Initial and the typical distribution of the number of servers in the five operating regions; average initial server load: $30\%$ row 1 and  $70\%$ row 2. (b) The ratio of in-cluster to local decisions in response to scaling requests versus time; average initial server load: $30\%$ row 3 and $70\%$ row 4.}
\label{ClusterSize20Load3070}
\end{figure*}

Figure \ref{ClusterSize20Load3070} shows that for a small cluster size, $n_{\mathcal{C}}=20$, the initial and the typical distribution of the number of servers in the five operating regions depends on the load. As expected, at low load (average load $30\%$ of the server capacity), the initial server distribution is concentrated in operating regions at the left and in the optimal region $\mathcal{R}_{3}$: $15\%$ of the servers operate in $\mathcal{R}_{1}$, $40\%$ in $\mathcal{R}_{2}$, $35\%$ in $\mathcal{R}_{3}$, $10\%$ in $\mathcal{R}_{4}$, and there are no servers in $\mathcal{R}_{5}$. During  the typical operation the majority, $65\%$ are in $\mathcal{R}_{3}$ (optimal), $25\%$ and $5\%$ are in the suboptimal low and high, $\mathcal{R}_{2}$ and $\mathcal{R}_{4}$, respectively, and none in the undesirable high.
On the other hand, at high load (average load $70\%$ of the server capacity) the initial server distribution is concentrated in operating regions at the right and in the optimal region; no servers operate in $\mathcal{R}_{1}$, $5\%$ in $\mathcal{R}_{2}$, $30\%$ in $\mathcal{R}_{3}$, $40\%$ in $\mathcal{R}_{4}$, and $25\%$ in $\mathcal{R}_{5}$. During the typical operating mode none of the servers operate in $\mathcal{R}_{1}$, $5\%$ in $\mathcal{R}_{2}$, $60\%$ in $\mathcal{R}_{3}$, $30\%$ in $\mathcal{R}_{4}$, and $5\%$ in $\mathcal{R}_{5}$.

\begin{figure*}[!ht]
\begin{center}
\includegraphics[width=7.5cm]{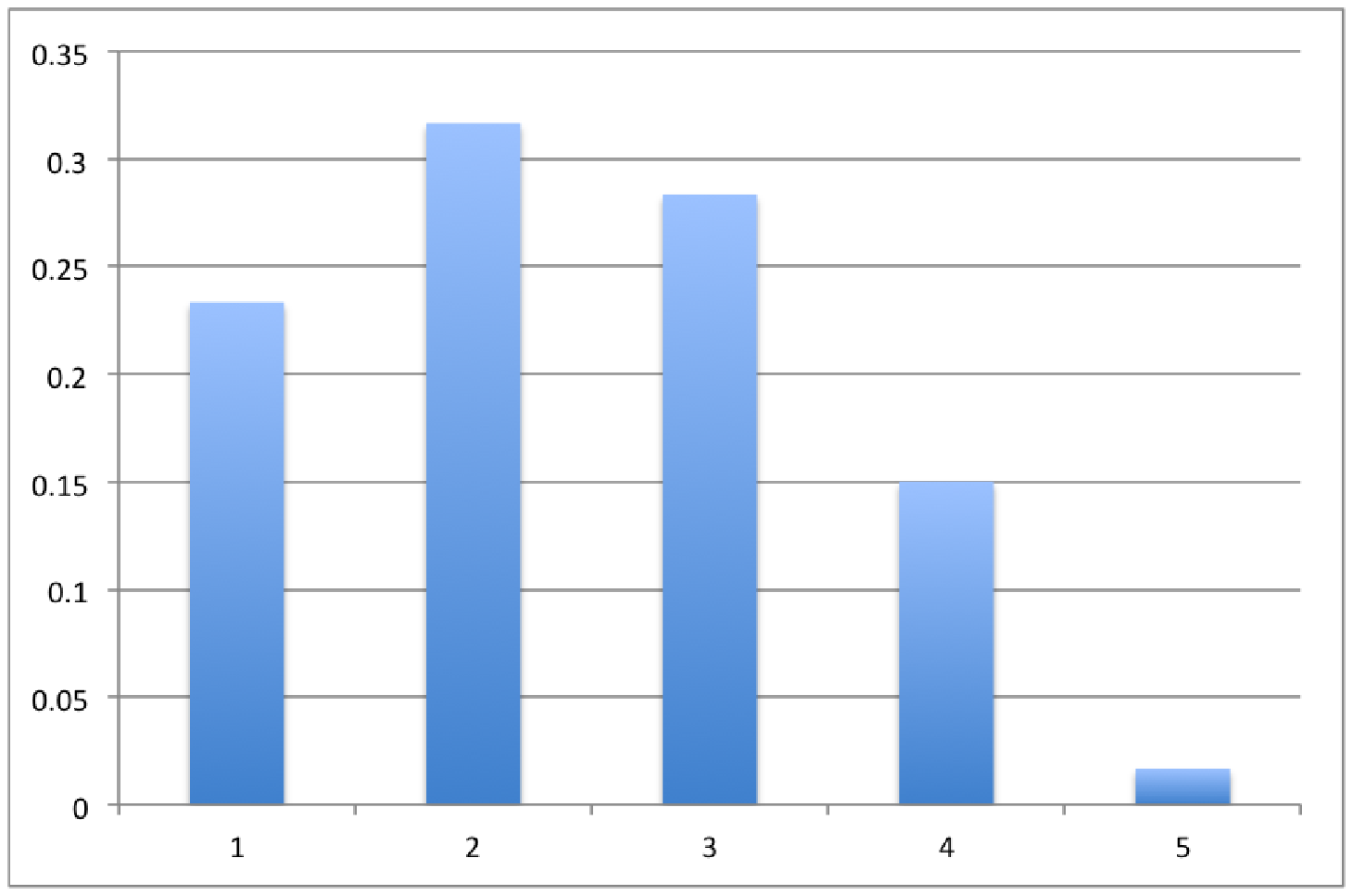}
\includegraphics[width=7.5cm]{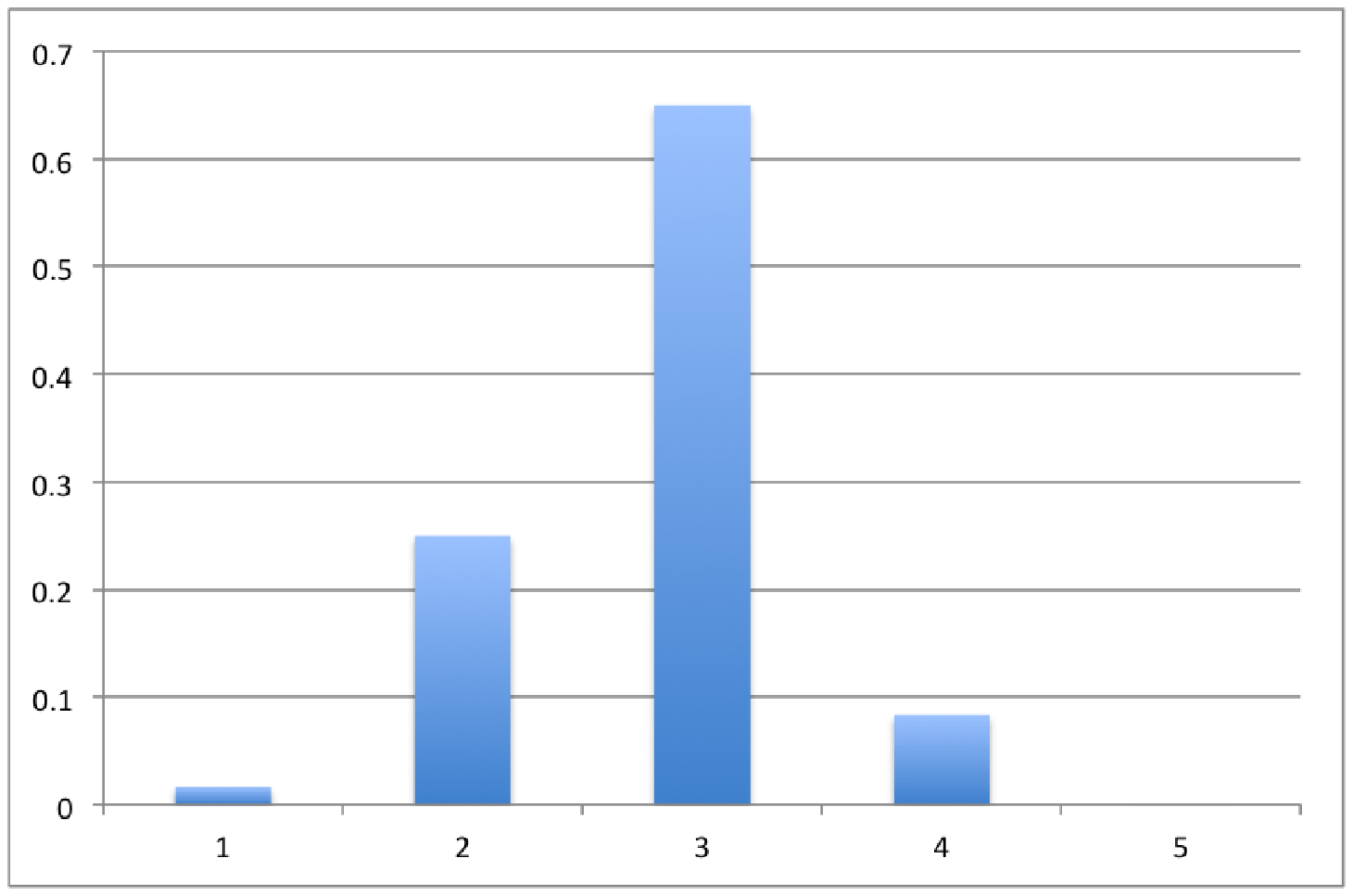}\\
\includegraphics[width=7.5cm]{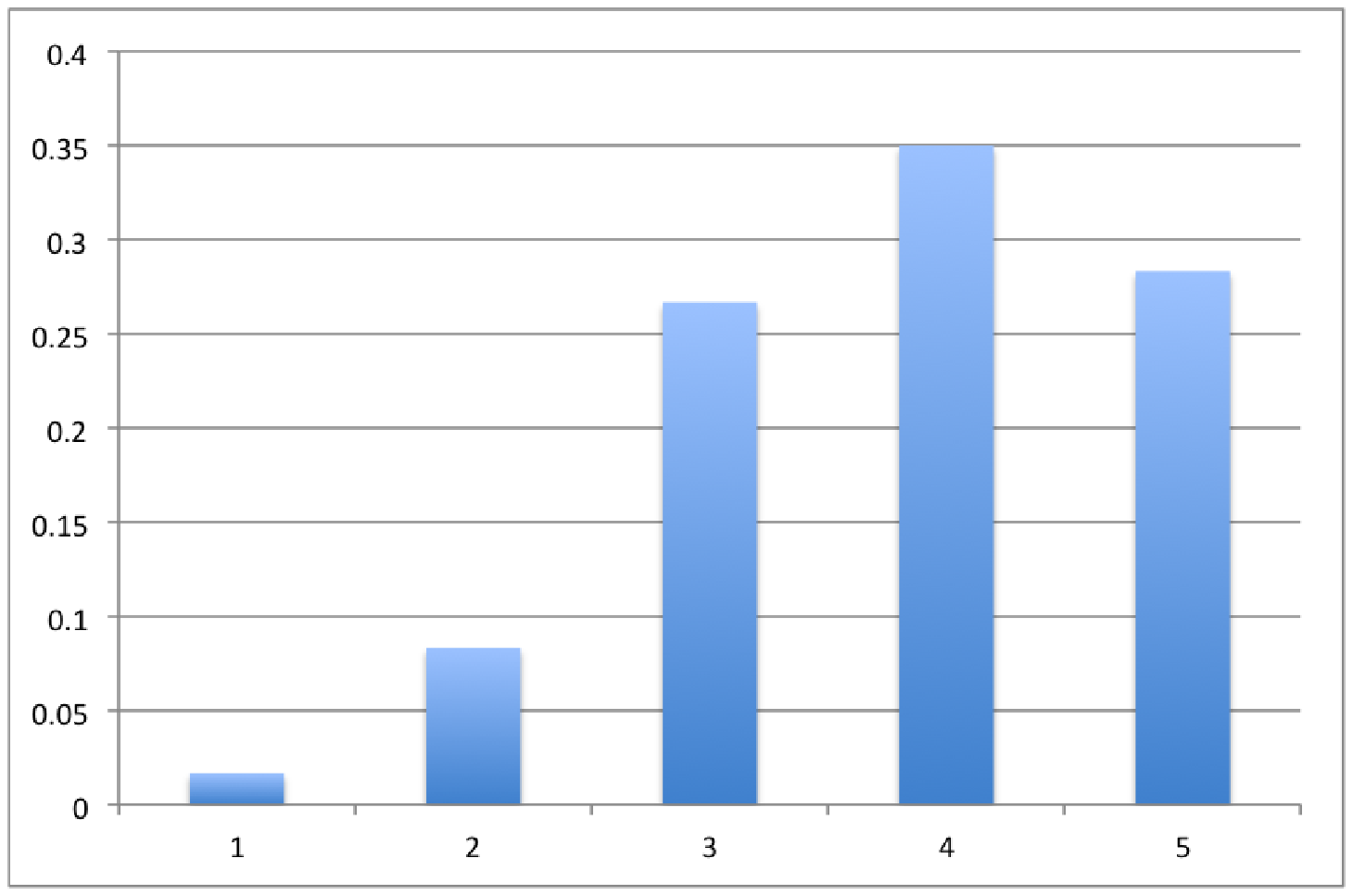}
\includegraphics[width=7.5cm]{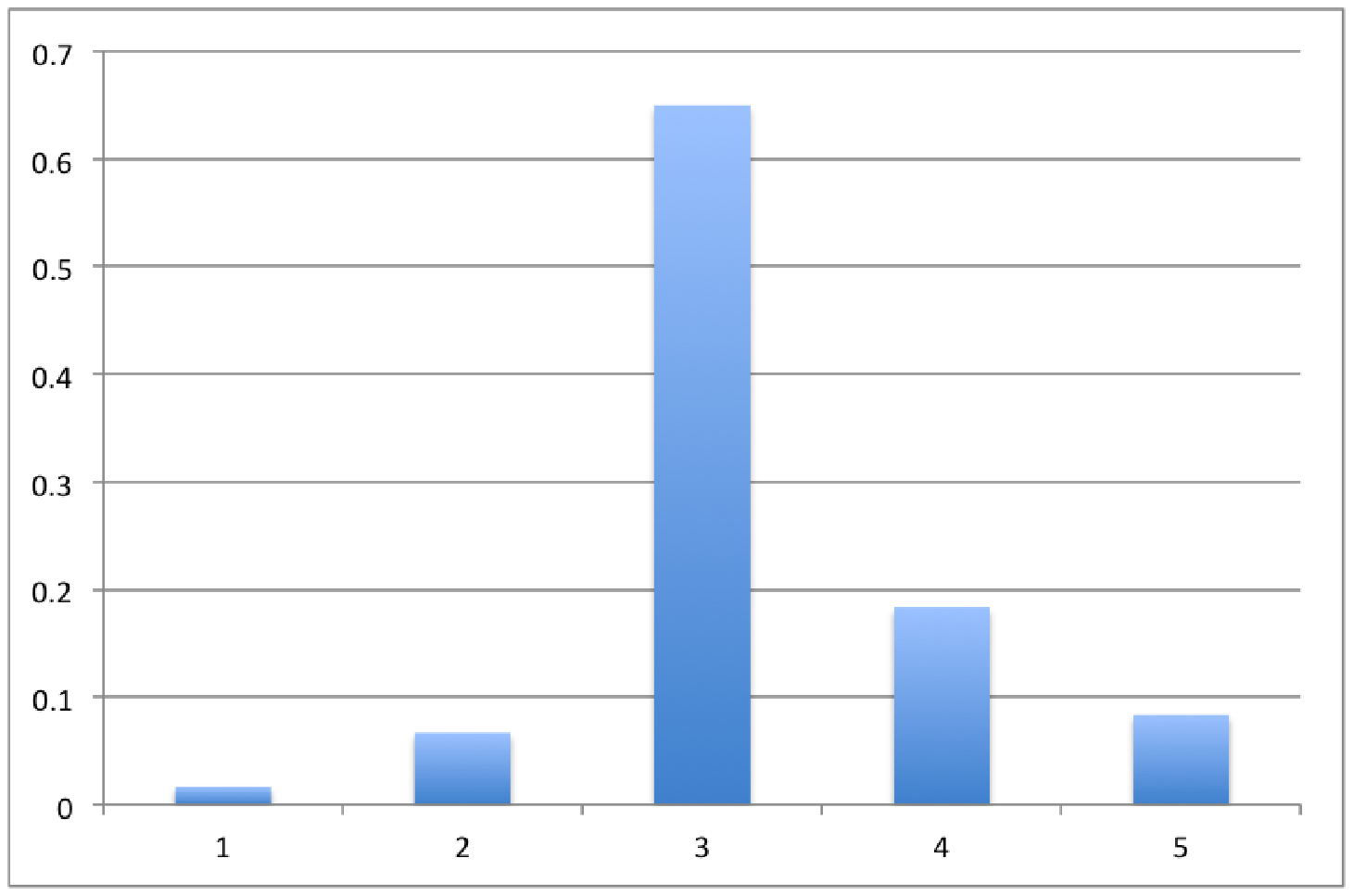}\\
~~~~~~~~~~Initial~~~~~~~~~~~~~~~~~~~~~~~~~~~~~~~~~~~~~~~~~~~~~~~Typical~~~~~~\\
\includegraphics[width=12cm]{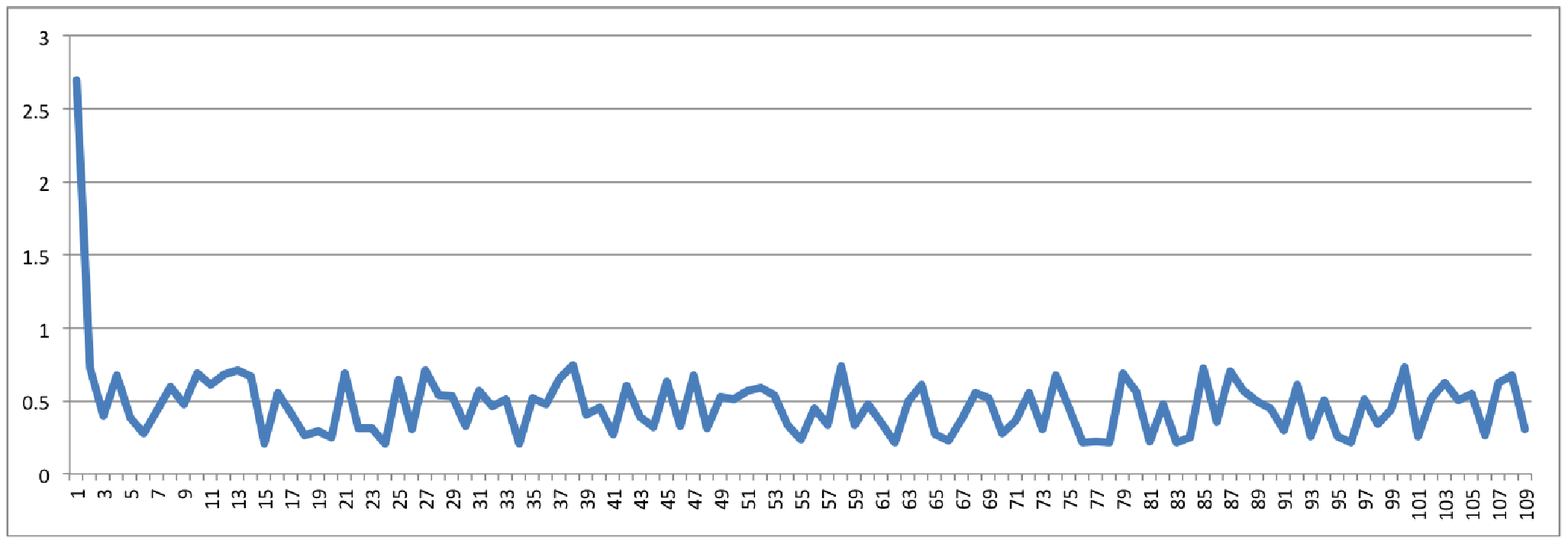}
\includegraphics[width=12cm]{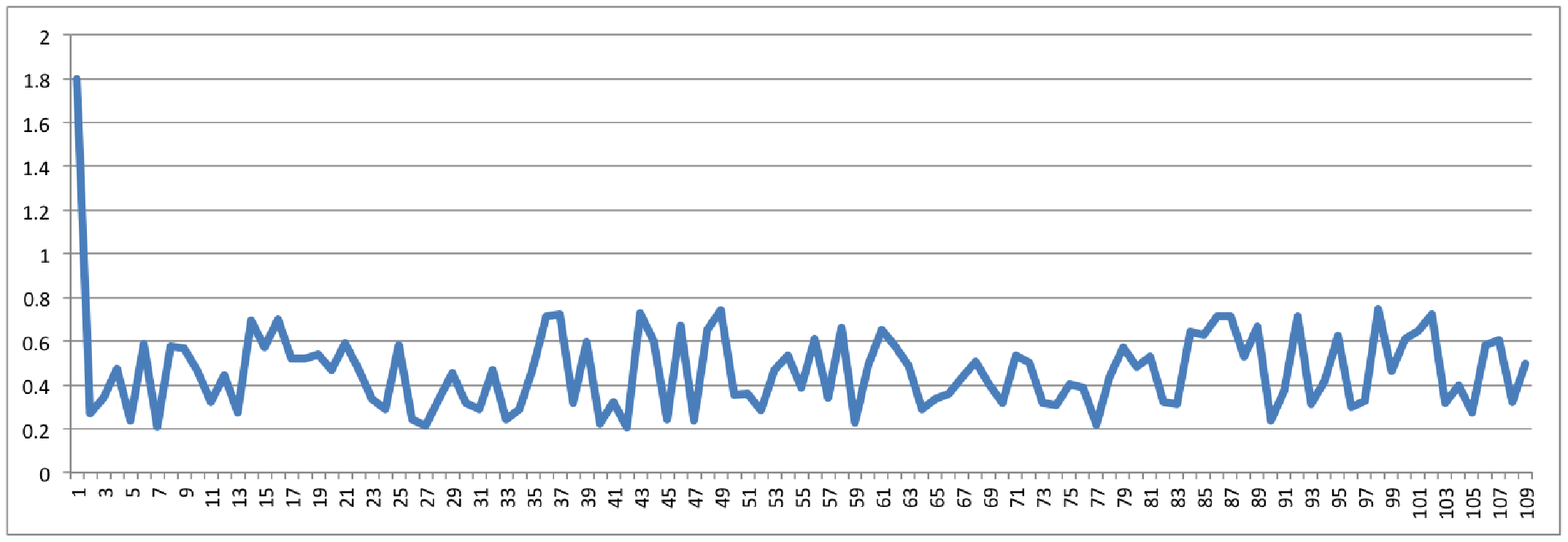}
\end{center}
\caption{The effect of the initial server load when there are 60 servers in the cluster. (a) Initial and typical distribution of the clusters in the five operating regions; average initial server load: $30\%$ row 1 and  $70\%$ row 2. (b) The ratio of in-cluster to local decisions in response to scaling requests versus time; average initial server load: $30\%$ row 3 and $70\%$ row 4.}
\label{ClusterSize60Load3070}
\end{figure*}

\begin{figure*}[!ht]
\begin{center}
\includegraphics[width=7.5cm]{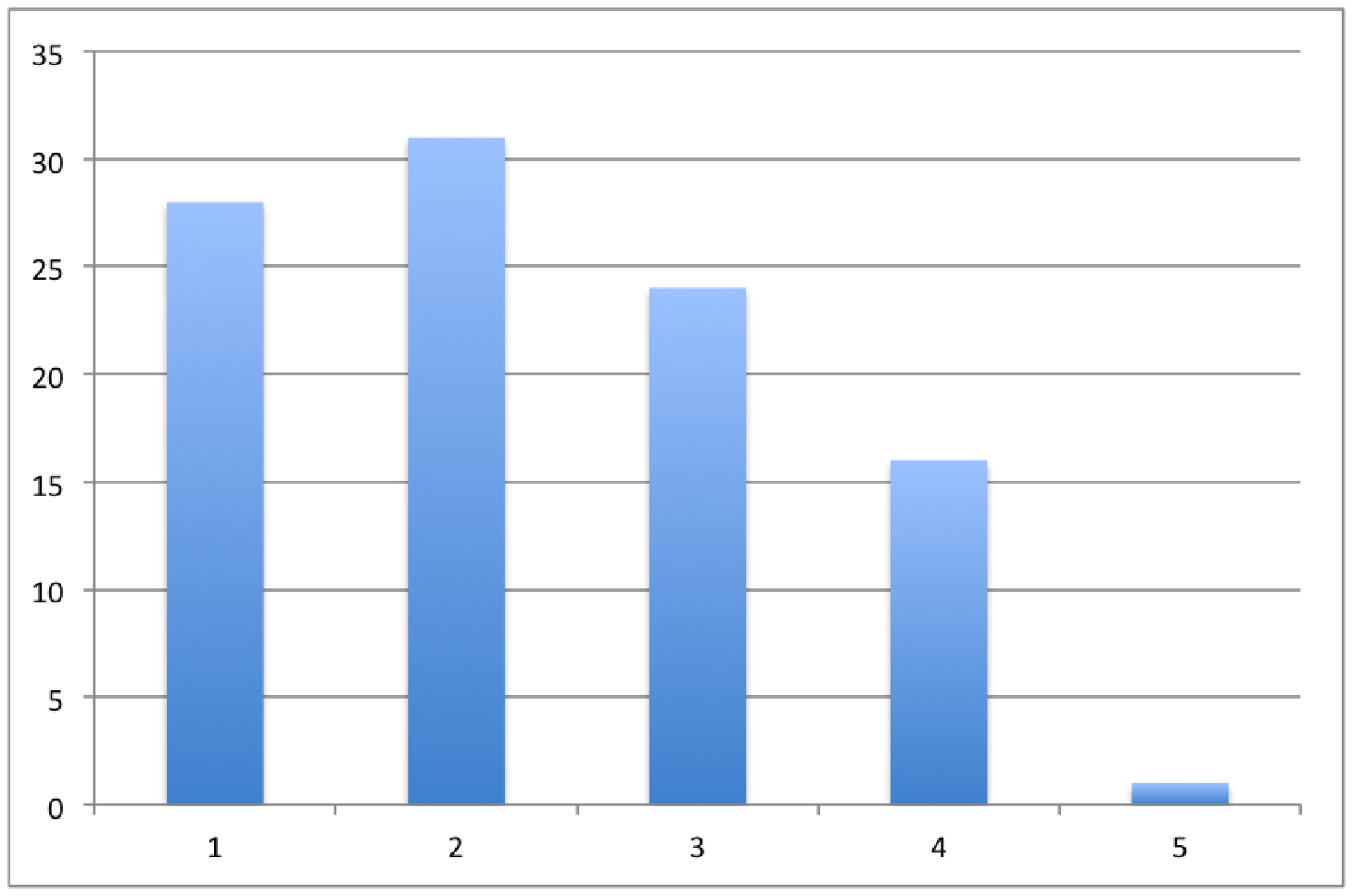}
\includegraphics[width=7.5cm]{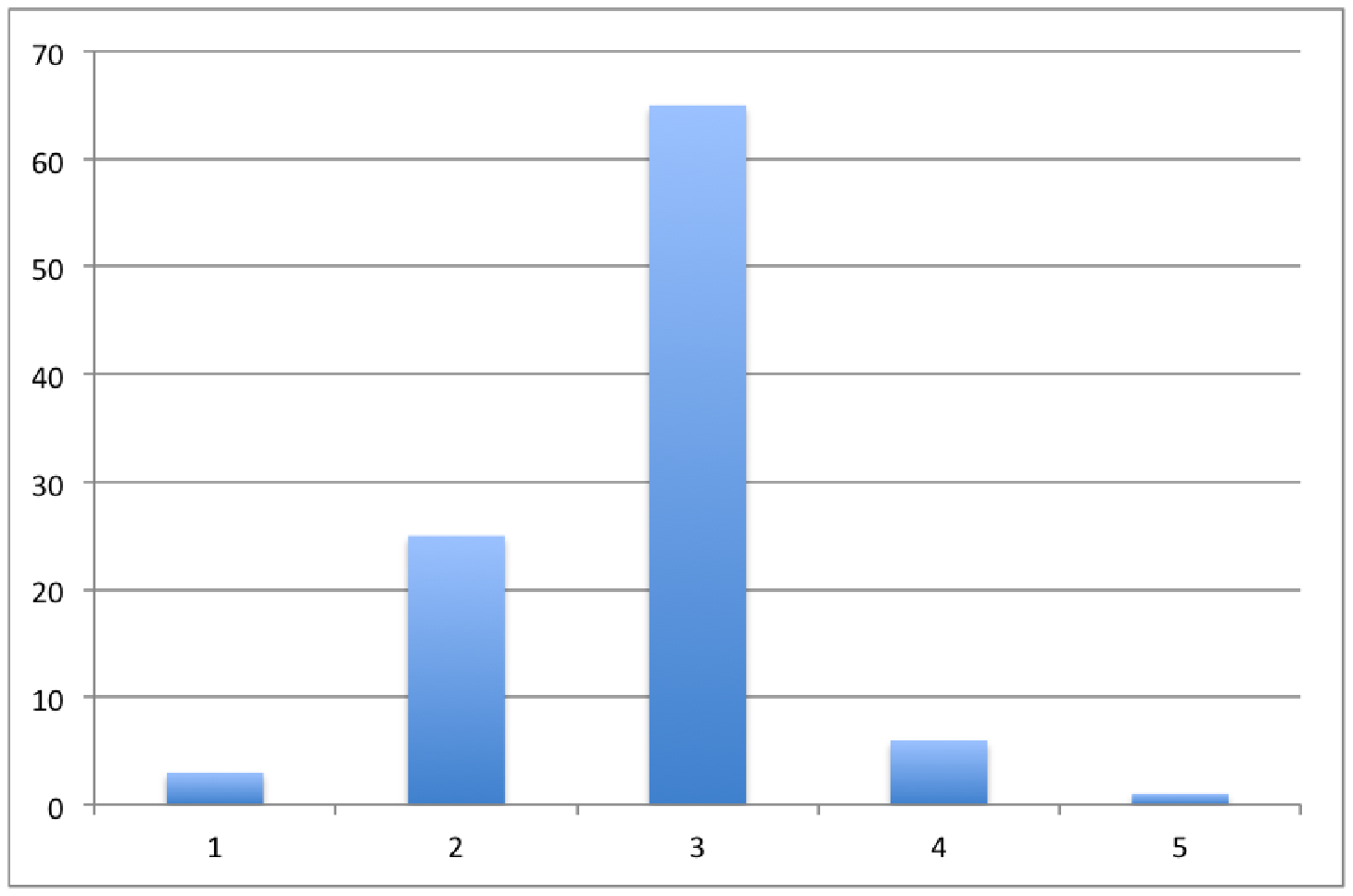}\\
\includegraphics[width=7.5cm]{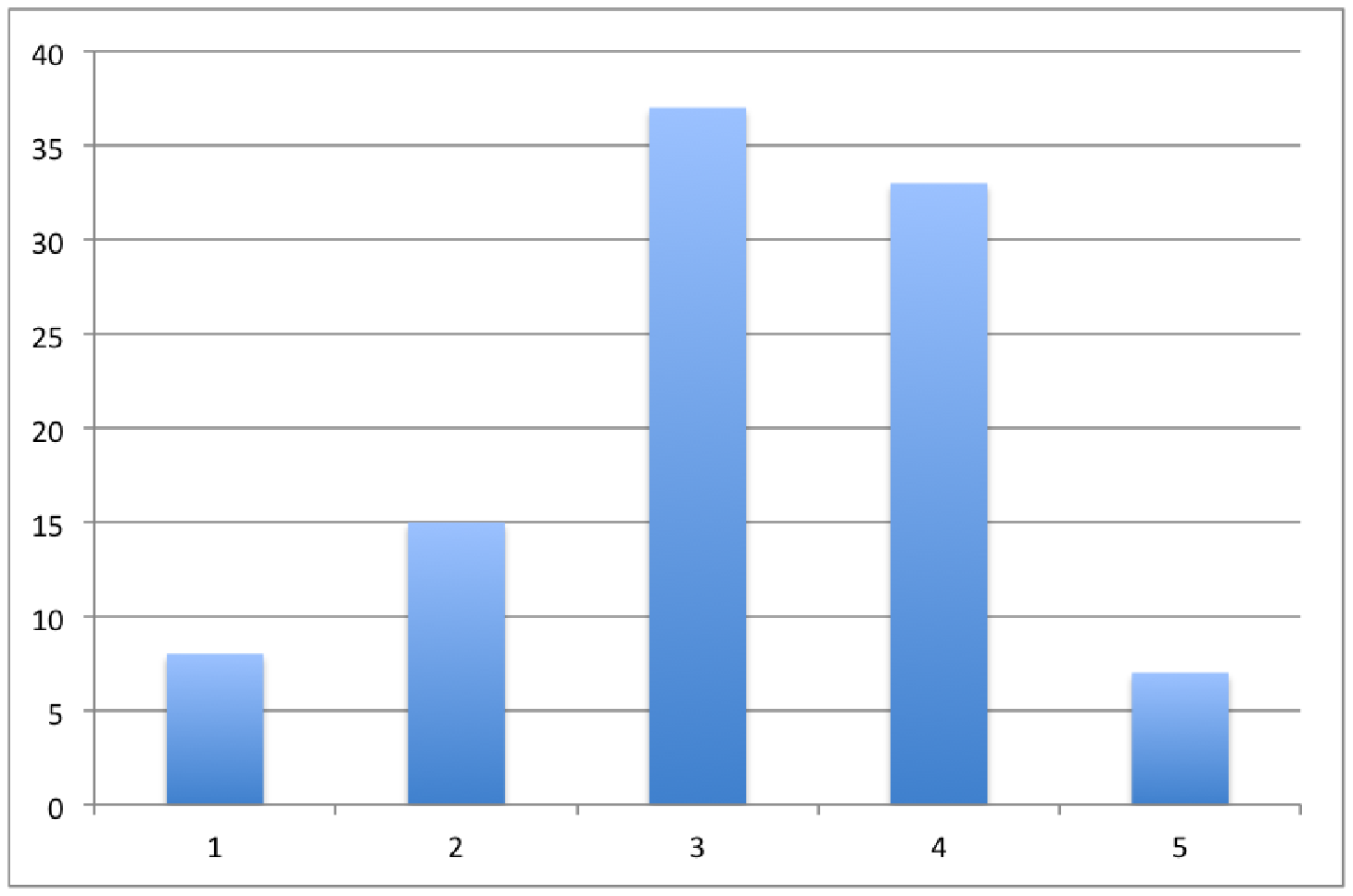}
\includegraphics[width=7.5cm]{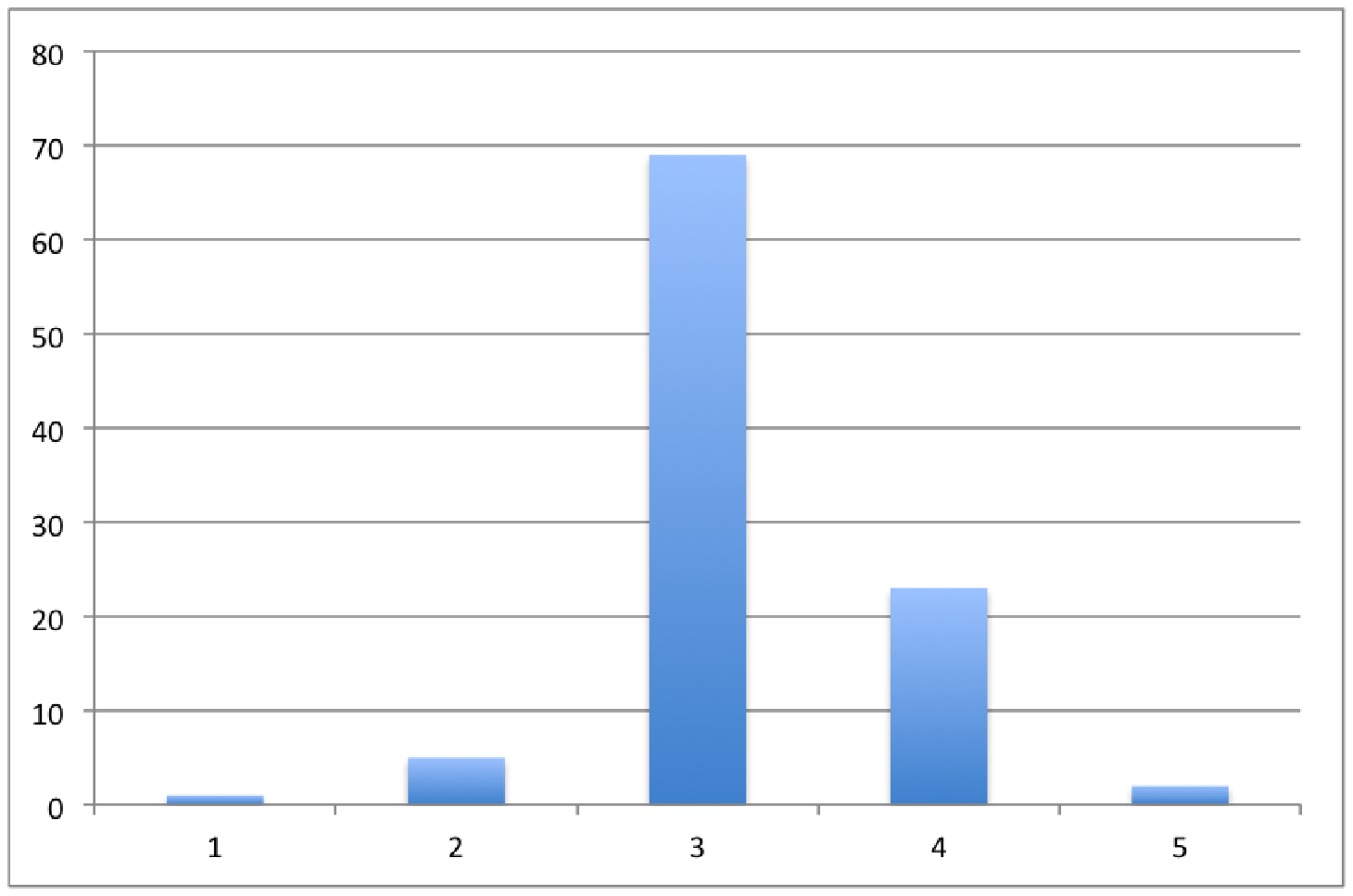}\\
~~~~~~~Initial~~~~~~~~~~~~~~~~~~~~~~~~~~~~~~~~~~~~~~~~~~~~~~~Typical~~~~~~\\
\includegraphics[width=12cm]{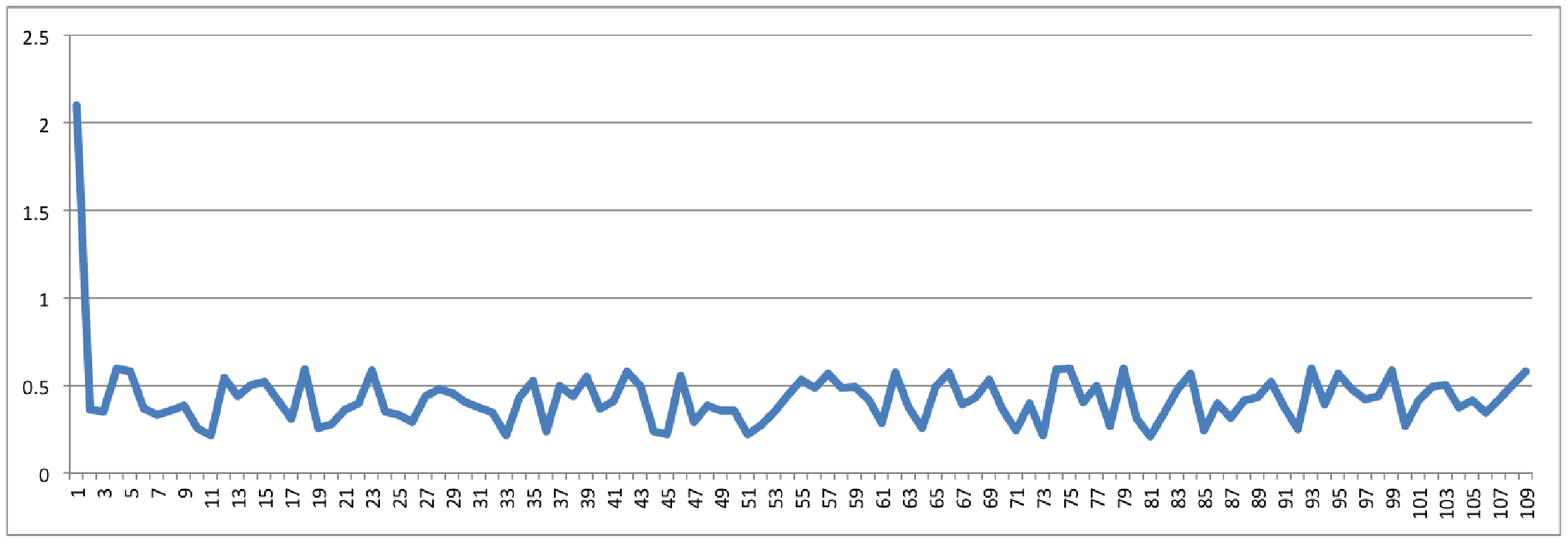}
\includegraphics[width=12cm]{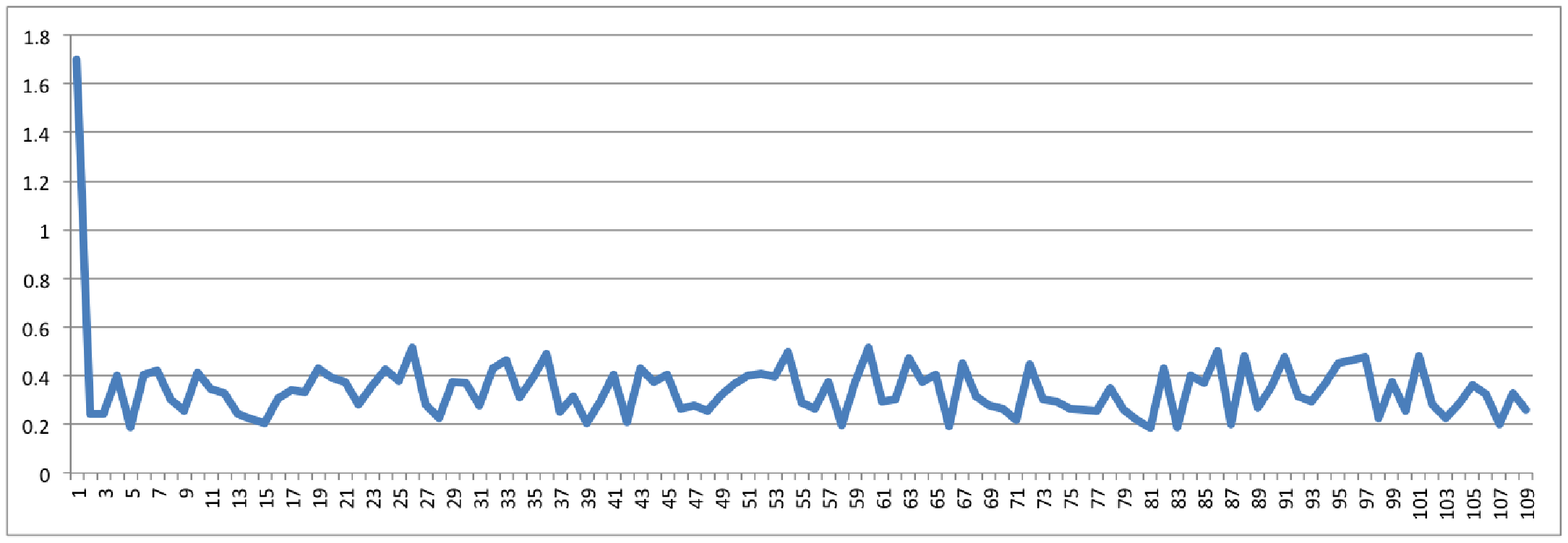}
\end{center}
\caption{The effect of the initial server load when there are 100 servers in the cluster. (a) Initial and the typical distribution of the clusters in the five operating regions;average initial server load: $30\%$ row 1 and  $70\%$ row 2. (b) The ratio of in-cluster to local decisions in response to scaling requests versus time; average initial server load: $30\%$ row 3 and $70\%$ row 4.}
\label{ClusterSize1000Load3070}
\end{figure*}

We observe a similar behavior in Figures \ref{ClusterSize60Load3070} and \ref{ClusterSize1000Load3070} for cluster sizes  $n_{\mathcal{C}}=60$ and $n_{\mathcal{C}}=100$, respectively. We conclude that in all cases during the typical operation the fraction of servers in the optimal region is in the $65-75\%$ range regardless of the load. The three figures show that the typical ratio of in-cluster to local decisions is qualitatively similar and that most scaling decisions are made locally, without the interactions with the cluster leader, regardless of the load placed on the system.

\section{Conclusions \& Future Work}
\label{Conclusions}
\medskip

Cloud elasticity is one of the most important reasons why the cloud computing has attracted so many users and organizations in such a short period of time. Elasticity means that an application can get as much resources as it needs at any given time. But it is economically unfeasible to support cloud elasticity solely by over-provisioning, in other words by guaranteeing that the available cloud computing center capacity can accommodate a peak load many times larger than the average  load.

One of the reasons why over-provisioning cannot be sustained is that the energy footprint of cloud computing centers cannot grow at the current rate. Increasing the energy efficiency of individual components of a server, of the communication infrastructure, and of the cooling systems will contribute to a lower energy footprint. At the same time, it seems obvious that energy-aware cloud resource management policies have to enforced and used to guarantee that existing commitments are satisfied, e.g., running applications are able to scale, and that more accurate information about the sate of the cloud should be available to support admission control, capacity allocation, and QoS for different classes of applications.

In this paper we consider a clustered organization and develop energy-aware algorithms for application scaling and resource management in a cluster. We recognize five operating regions of a server, as shown in Figure \ref{ServerPerformanceVsEnergy} which displays  the normalized performance of server $\mathcal{S}_{k}$ function of the power level: optimal, $\mathcal{R}_{3}$, suboptimal low and high, $\mathcal{R}_{2}$ and $\mathcal{R}_{4}$, respectively, and undesirable low and high $\mathcal{R}_{1}$ and $\mathcal{R}_{5}$,  respectively. The objective of the algorithms is to ensure that the largest possible number of running servers operate within the boundaries of their respective optimal operating regions.
Another critical policy is to migrate applications from a lightly loaded server and then to switch the server to a sleep state and avoid at all costs to keep a server in an idle state when it consumes a significant amount of power. 

The model we developed is quite general. Though targeting primarily public clouds and the  {\it IaaS} delivery model the models and algorithms introduced in this paper can be applied to private and hybrid clouds and to the {\it SaaS} and the {\it PaaS} cloud delivery models. We have not discussed the case of processors using the Dynamic Voltage and Frequency Scaling (DVFS) techniques but the algorithms apply to such cases as well; indeed, instead of static parameters delimiting the five operating regions these parameter change with voltage and frequency. The only restriction is that we limit the scaling rate of an application in each evaluation interval, thus a system may not be able to accommodate ``flash events,'' sudden drastic increases of the system load.

The simulation results discussed in Section \ref{SimulationExperiment} show that the algorithm is able to increase the number of servers operating in region $\mathcal{R}_{3}$ and decrease the number of those operating in the two undesirable regions. The effect of  the algorithm is that in a typical state  about $70\%$ of servers end up operating in the optimal region and only  about $5\%$ of them are in the two undesirable regions, leaving about $25\%$ in the two suboptimal regions. The simulation experiments show that these results are valid for a fair range of the number of servers in a cluster, from $20$ to $100$.

We also investigated the effectiveness of the algorithm for different average server load, from light load, average server load $30\%$, medium server load, $50\%$, and high load, $70\%$, of their capacity. We see that there are only small variations of the results in all these cases thus, we concluded that the algorithm is invariant to the system load. 

Figure \ref{ServersVsRegion} shows that throughout the simulation the vast majority of servers operate in the optimal region. It also shows that whenever possible the algorithms switches servers to a sleep mode. For example, during the observation intervals labeled $42, 50$ and $76$ only $27, 25$ and $25$, respectively, out of the $40$ servers are running, the other $13 - 15$ are in a sleep mode. We also see that throughout the simulation the number of servers in undesirable regions is very low. These results show that the algorithm is very effective in forcing servers at, or near to, optimal performance to energy consumption operating points.

The amount of energy saved during an interval of time $[t_{s}, t_{e}]$ by a cluster $\mathcal{C}$ 

\begin{equation}
E_{saved}^{[t_{s}, t_{e}]} = E_{0}^{[t_{s}, t_{e}]} \left[ 1 - {\bar{\xi}_{\mathcal{C}}^{0}  \over \bar {\xi}_{\mathcal{C}}^{ea}}  \right].
\end{equation}
with

$ E_{0}^{[t_{s}, t_{e}]}$ -the energy used when the cluster operates without the energy-aware algorithm support,

$\bar{\xi}_{\mathcal{C}}^{0}$ - the efficiency (see Equation \ref{AverageClusterEfficiency}) when the cluster operates without the energy-aware algorithm support

$\bar{\xi}_{\mathcal{C}}^{ea}$ - the efficiency when the cluster operates with the energy-aware algorithm support; the cluster is expected to deliver the same level of performance with and without the the energy-aware algorithm support.

We expect that the algorithm will have its most significant impact on a lightly loaded system because a fair number of lightly loaded servers will be switched to a sleep state thus, saving a fair amount of energy.

There is always a price to pay for an additional function provided by a system, so we have to evaluate the overhead of the algorithm. We assume a clustered organization of the cloud, a set of servers managed by a cluster leader which implements local resource allocation policies; the cluster leaders communicate for implementing global policies. We want as accurate state information as feasible and the alternative, a centralized control in a cloud with several million servers cannot possibly allow the management center to have accurate state information, this is physically unfeasible due to large communication delays and contention for access to the management systems.

The computational overhead of the algorithm is minimal thus, we are primarily concerned with its communication complexity. The simulation experiments show that after the initial transient period when most decisions require the interventions of the leader, there is a balance between local and non-local decisions. All our experiments show that, with the exception of brief periods of time, as many as twice as many decisions are made locally, thus there is no communication with the cluster leader. The algorithm seems to exploit a form of {\it locality},  many scaling decisions are made locally, without the involvement of the cluster leader.

We plan to investigate further optimizations of the algorithm and to implement a test bed system on our local clusters using one of the open cloud platforms \cite{Marinescu13}. We also plan to investigate the integration of the algorithm in {\it Xen}.

\bigskip

{\it Ashkan Paya.} Ashkan Paya is a second year graduate student in the EECS Department at University of Central Florida pursuing a Ph.D. degree in Computer Science. He graduated from Sharif University of Technology in Teheran, Iran, with a BS Degree in Computer Science in 2011. His research interests are in the area of resource management in large-scale systems and in cloud computing

\bigskip

{\it Dan C. Marinescu.} In 1984 Dan Marinescu joined the Computer Science Department at Purdue University in West Lafayette, Indiana as an Associate and the Full Professor. Since August 2001 he is a Provost Research Professor and Professor of Computer Science at University of Central Florida. His research interests are: scientific computing, process coordination and distributed computing including cloud computing and quantum information processing. He has published more than 210 papers in referred journals and conference proceedings. He published several books: {\it Internet-based Workflow Management} published by Wiley in 2002, {\it Approaching Quantum Computing} (co-authored with Gabriela M. Marinescu, Prentice Hall - 2005); {\it Classical and Quantum Information}, (co-authored with Gabriela M. Marinescu) published in February 2011 by Academic Press, a division of Elsevier, and {\it Cloud Computing: Theory and Practice} published by Morgan Kaufmann in 2013.

\end{document}